\documentclass{emulateapj}
\usepackage{lscape}

\begin{document}

%\slugcomment{ Feb 23, 2006}

\shorttitle{MSX cores}
\shortauthors{Ragan et al.}

\title{Molecular Line Observations of Infrared Dark Clouds: 
Seeking the Precursors to Intermediate and Massive Star Formation}

\author{S.~E.~Ragan\altaffilmark{1}, E.~A.~Bergin\altaffilmark{1}, 
R.~Plume\altaffilmark{2}, D.~L.~Gibson\altaffilmark{2}, 
D.~J.~Wilner\altaffilmark{3}, S.~O'Brien\altaffilmark{4} \& E.~Hails\altaffilmark{5}}
%\affil{}
%\email{}

\altaffiltext{1}{Dept. of Astronomy, University of Michigan, Ann Arbor, MI, USA}
\altaffiltext{2}{Dept. of Physics and Astronomy, University of Calgary, Calgary, AB, Canada}
\altaffiltext{3}{Harvard-Smithsonian Center for Astrophysics, Cambridge, MA, USA}
\altaffiltext{4}{Dept. of Physics, University of Notre Dame, Notre Dame, IN, USA}
\altaffiltext{5}{Arizona Radio Observatory, Tuson, AZ, USA}

\begin{abstract}
We have identified 41 infrared dark clouds from the 8 $\mu$m maps of the 
Midcourse Space Experiment (MSX), selected to be found within one square 
degree areas centered on known ultracompact HII regions.
We have mapped these infrared dark clouds in 
N$_{2}$H$^{+}$ $1 \rightarrow 0$, CS $2 \rightarrow 1$ 
and C$^{18}$O $1 \rightarrow 0$ emission using the Five College Radio 
Astronomy Observatory. The maps of the different species often show 
striking differences in morphologies, indicating differences 
in evolutionary state and/or the presence of undetected, deeply embedded 
protostars. 
We derive an average mass for these clouds using N$_2$H$^+$ column densities 
of $\approx 2500$ M$_{\odot}$, a value comparable to that found in previous
studies of high mass star forming cores using other mass tracers.
The linewidths of these clouds are typically $\sim$2.0 - 2.9 km~s$^{-1}$.
Based on the fact that they are dark at 8 $\mu$m, compact, massive, and 
have large velocity dispersions, we suggest that these clouds may be 
the precursor sites of intermediate and high mass star formation.  
\end{abstract}


\section{Introduction} 
\label{intro}
The study of star formation has made tremendous strides over the past two
decades.  Advances in observational capabilities have allowed a number of 
phases of the star formation process to be identified and characterized, 
starting with the centrally concentrated core of molecular gas that collapses 
to form a star surrounded by a proto-planetary disk.  It has been the isolation
of such objects that have not yet formed stars -- pre-stellar cores -- that 
has allowed us to probe the earliest initial stages of star formation 
(see Andr\'{e} et al. 2000; Alves, Lada \& Lada 2000).  
Most of the progress has focused on the formation of low-mass stars, 
predominantly because these objects can form isolated from other nearby stars 
that reduces confusion, and, conveniently, there is a large sample of such 
objects located in nearby clouds. 
However, it has now been recognized that most stars that are more massive than 
the Sun do not form in isolated fashion, but rather in clusters of $>$100 stars
(Zinnecker et al. 1993).
 
Progress in our understanding of high-mass star formation has been hampered by 
a number of factors:
(1) timescales for massive star formation are short and examples in a given
state are intrinsically rare,
(2) the large distances to the Giant Molecular Clouds (GMCs) that are the 
birth sites of massive stars make studies of individual objects difficult, 
and (3) the very nature of clustered star formation increases confusion 
(cf. Garay \& Lizano 1999).  As a result, the objects in the earliest phases --
the ``pre-stellar massive cores''  -- have been difficult to identify.  

The traditional method of locating low-mass cores has been to examine optical 
plates for regions of obscured starlight and then to pursue follow up molecular
line observations (Myers \& Benson 1983; Lee \& Myers 1999).  A comparison with
the IRAS point source catalog then denotes whether these cores are associated 
with newly formed stars (Beichmann et al. 1986).  This method cannot be applied
to massive star forming regions since the greater distances makes isolating 
individual objects difficult.  Moreover, the size and high column densities of 
GMCs makes it impossible to use optical plates to find individual objects.  
An analogous method of searching for molecular cores is to search for obscured 
regions in galactic mid-infrared background. However, due to atmospheric 
constraints, ground-based observations in the mid-IR are difficult to obtain. 
The ISOCAM instrument on the Infrared Space Observatory was used in this 
fashion, but only with pointed observations towards previously identified 
cores (e.g. Bacmann et al. 2000). 

The Midcourse Space Experiment (MSX) surveyed the galactic plane in 
mid-infrared bands spanning from 7 to 25~$\mu$m. This survey revealed a large 
population of dark clouds, predominantly located toward the inner galaxy 
(Egan et al. 1998).  Follow-up molecular studies of a few objects confirmed 
that the obscured regions represent a new population of dense, 
n(H$_2$) $> 10^5$ cm$^{-3}$, and cold, T $< 20$ K, molecular clouds 
(Carey et al 1998).  Further comparison to IRAS images demonstrated that most 
of these clouds are dark from 7--100~$\mu$m, presumably because these objects 
either do not contain newly formed stars, or any newly formed stars are 
very deeply embedded.  As such, a sub-sample of these objects may trace massive
pre-stellar cores.  There has been substantial activity in this field recently,
with numerous groups analyzing various samples of infrared dark clouds (IRDCs). 
These studies have shown that it is likely that IRDCs are the birth-sites of 
high-mass stars and stellar clusters (Sridharan et al. 2005, 
Menten et al. 2005, Rathborne et al. 2006, Pillai et al. 2006).

We have identified a sample of infrared dark clouds and searched them for 
emission from the N$_2$H$^+$ $1 \rightarrow 0$, CS $2 \rightarrow 1$ 
and C$^{18}$O $1 \rightarrow 0$ transitions ($\S$\ref{obs}). 
In most cases, we find that the emission closely corresponds to the MSX 
dark regions.  Using a gas temperature of T=15~K based on 
CO $1 \rightarrow 0$ data, we deduce several properties of the 
dark clouds including column density and mass ($\S$\ref{results}).  
We summarize our findings and the implications ($\S$\ref{summary}).

\section{Source Selection \& Observations}
\label{obs}
To search for pre-stellar massive cores we have compiled a catalog of 
MSX dark clouds. This catalog is biased, as we have only searched the 
released MSX band A (centered at 8.8 $\mu$m) images for infrared dark clouds
in the vicinity of known ultra-compact (UC) HII regions from the 
Wood \& Churchwell (1989) catalog.  More specifically, we searched for 
absorbing clouds within a square degree centered on a given UC HII region.  
This strategy takes advantage of the fact that young stars generally form 
in clusters, and, therefore, a good place to search for the pre-cursors to 
massive stars is in the vicinity of regions with current massive star 
formation evidenced by the UC HII regions.
In this fashion we have isolated 114 infrared dark clouds, of which only a 
small fraction (15\%) have known associations with radio sources or masers.  
From this catalog of 114 infrared dark clouds we selected a sub-sample of 
the 41 most compact and most opaque ($\tau_{8.8\mu m}$ $\gtrsim$ 0.4) 
cores and targeted these for molecular line observations.   

We mapped 41 of the sample of MSX dark clouds 
in emission from 
C$^{18}$O $J = 1 \rightarrow 0$ ($\nu = 109.782$ GHz), 
CS $J = 2 \rightarrow 1$ ($\nu = 97.981$ GHz), and 
N$_2$H$^+$ $J = 1 \rightarrow 0$ ($\nu = 93.173$ GHz) 
using the 14m Five College Radio Astronomy Observatory (FCRAO).  
The observations were made in 2002 February, May \& December using the 
16 element focal plane array receiver SEQUOIA.  Each $2.5\arcmin \times 2.5 \arcmin$ region was mapped with the 50$\arcsec$ beam, with typical rms noise levels of $\sim0.05 - 0.1$ K.
We used the Narrow Band Correlator backend configured to a velocity resolution 
of $\sim 0.13$ km~s$^{-1}$.  Typical system temperatures (T$_{sys}$) were 
$200 - 300 K$. 
Main beam efficiencies ($\eta_{mb}$) were approximated at 50\% from the 
standard FCRAO values. This is accurate within a few percent for each 
transition.  For each spectrum, a first-order, linear baseline was fit to 
remove instrumental and continuum offsets and drift.  

This selection of species is motivated by studies of low-mass pre-stellar 
clouds. N$_2$H$^+$ is unlikely to significantly suffer from the effects of 
depletion as the core condenses, and this species is a good tracer of the dense 
centers of starless cores (Bergin \& Langer 1997; Tafalla et al. 2002).  
Conversely, C$^{18}$O and CS emission can be used to trace the outer layers.
Table~1 lists some basic dark cloud properties such as the HII region name, 
the coordinates, the size, and the center-to-edge brightness contrast.
The brightness contrast was obtained by comparing the brightness in the band 
centered at 8.8 $\mu$m at the center of the core with the average background 
brightness, estimated from an average of the intensities in a vertical and 
horizontal slit across the dark region. If the dark cloud was searched for 
molecular emission, we provide the LSR velocity range to which the observations 
were sensitive.

In an appendix, we provide a source by source description of any associations with known star formation indicators (e.g. IRAS sources, masers, radio continuum sources). The large majority of our sources have no association with any known strong infrared source.

\section{Results} 
\label{results}

\subsection{Molecular Line Fits} 
\label{linefits}
Each line is fit with a Gaussian profile to determine the integrated 
intensity, line width, and the LSR velocity of the emitting material.  Table~2 presents the results of the spectral line fitting.  All parameters were extracted by standard Gaussian fitting methods in the 
CLASS package (Buisson et al. 2002); for N$_2$H$^+$, the seven main hyperfine 
components were fit together using the HFS routine.  The reported line-center velocity corresponds to that of the strongest hyperfine component ($J=1 \rightarrow 0$, $F_{1},F = 2, 3 \rightarrow 1, 2$) at 93.1738~GHz.  

Our observations were obtained with velocity resolution of 0.13 km~s$^{-1}$, and to increase the signal to noise, we on occasion  smoothed the line profiles by a factor of 2.  As such, we had little sensitivity to structure within the line.  Within our sensitivity limits (see $\S${\ref{obs}}), we see no evidence for extended line wings, and all lines were well-fit by single Gaussians.  We report here only basic line properties.

\subsection{Molecular Emission Morphologies} 
\label{morphologies}
Figure~1 displays the Midcourse Space Experiment (MSX) images of each dark
cloud overlayed with the integrated intensity maps of molecular emission for 
all observed molecular transitions.  The contour levels for each dark cloud are specified in the captions.  In some cases, multiple velocity components are detected, and the different velocity components are given different colors.  The blue contours always correspond to the molecular emission morphology that most closely corresponds to the distribution of 8~$\micron$ absorption seen in the MSX image, and, therefore, is most likely to be associated with the dark cloud.  The line properties of the other velocity components (i.e. emission that is unassociated with the absorbing cloud) are not reported.  There are three cases in which an a/b designation was assigned to distinguish two emission peaks.  For G14.33-0.57 and G23.48-0.53, there are two spatially distinct and separate emission peaks at approximately the same characteristic velocity.  This may suggest that in these cases, we were able to resolve spatially separated fragments of a cloud.
In the case of G37.44+0.14, there are two velocity components that appear to correspond to absorbing regions (the 40km~s$^{-1}$ feature corresponding to the central absorbing cloud; the 18km~s$^{-1}$ feature correpsonding to absorption to the north and west).  Based on our assumptions, this would indicate that there are two unassociated dark clouds apparent in the same region of the sky by chance.

The morphological differences seen in Figure~1 are striking.  Some maps show 
well defined cores in all three molecular tracers (e.g. G32.01+0.05).  Other maps, like G06.26-0.51, show fairly well defined cores in C$^{18}$O and CS, but nothing obvious is seen in the 
N$_2$H$^+$.  And others, like G09.21-0.22 , show a strong centrally 
concentrated N$_2$H$^+$ core but CS that is much more diffuse. 
Finally, there are cores like G10.59-0.31 that show no real evidence for molecular emission centered on the dark cloud at all.  In this case, it is possible that any molecular emission directly associated with the dark cloud lies outside of the observed velocity band (see Table~1).  

These differences in the molecular emission maps may be the result of differences in the evolutionary states.  It is well known that CS and C$^{18}$O can form relatively quickly in the gas phase whereas N$_2$H$^+$ takes significantly longer.  However, as density enhances, the CS and C$^{18}$O tend to deplete onto the surfaces of dust grains, whereas N$_2$H$^+$ will remain in the gas phase.  Upon protostar formation, CS and C$^{18}$O can be released from dust grain surfaces (e.g. see models of Bergin \& Langer 1997; Lee et al. 2004).  Therefore, two scenarios can lead to the low abundance of N$_2$H$^+$ relative to CS or C$^{18}$O: the star forming core is at an early stage of condensation and the densities are low such that CS and C$^{18}$O would be not affected by dust depletion, or the presence of a protostar has released the CS and C$^{18}$O from dust grains, all the while the N$_2$H$^+$ abundance is essentially unchanged throughout the process.  Another possible explanation of the differences seen in the maps is that some of the dark clouds may contain as yet undetected protostars, obscured by the high opacity apparent in the 8.8~$\mu$m MSX images.  If a protostar is present, it can alter the local gas chemistry via grain mantle evaporation, which can change the emission morphologies (J{\o}rgensen 2004).  We are using data from the Spitzer Space Telescope to determine which dark clouds are truly starless and which contain embedded protostars in a sub-sample of these objects (Ragan et al. 2006, in prep.).

\subsection{Distance Estimates} 
\label{distances}
The kinematic distance to each dark cloud is calculated using using the line 
center velocity and the Milky Way rotation curve model of Fich, Blitz, \& Stark 
(1989).  The distance assignments are presented in Table~3 for dark clouds for which we estimated masses.  (Sources to which we are unable to assign a distance or those that show no significant emission are not subject to further calculations.)  For every position, there is both a ``near'' and ``far'' distance solution that corresponds to the characteristic velocity of the emission.  In addition to this ambiguity, Fich, Blitz, \& Stark (1989) cite a $\pm$14\% maximal deviation of the data from their rotation curve model;  based on this consideration, we calculate errors in the distances and provide them in Table~3.  In the cases where no error range (or an incomplete one) is given, no physical solution exists when calculating the distance with that error offset.  For all subsequent calculations, we assume that the dark cloud 
is located at the ``near'' distance.  We believe that this is a reasonable 
assumption since the clouds are seen in absorption against the Galactic 
mid-infrared background and, therefore, are unlikely to reside at the 
``far'' distance.  Assuming the ``near'' kinematical distance, which is also listed in Table~4, a typical core has a diameter of $\sim$0.9 pc.  

Interestingly, the kinematical distances for the velocity component associated with the absorption are seldom coincident with the distances estimated for the UC HII region that 
was the original search target.  In most cases we detect molecular emission at a single velocity, and no possible distance solution from the galactic rotation curve is consistent with the distance to the UC HII region.  However, in the cases where there are multiple velocity components, the distance to the UC HII region is often consistent with one of the kinematic distance solutions for a secondary velocity component.  For example, G37.44+0.14 has a secondary detection of a component at 18 km~s$^{-1}$, and the ``far'' distance associated with it (12.34 kpc) is very close to the distance to the UC HII region (12.0 kpc).  In this case, it is likely that we are detecting two clouds along the same line of sight at different distances: one near the UC HII region and one nearer to us.  Since the emission of the primary component corresponds so well to the absorbing dark cloud, we maintain that these lie at the ``near'' distances, though there is significant uncertainty in the distance calculation.
Nonetheless, as we will show, these clouds are massive and are likely associated with the formation of intermediate and high-mass stars and stellar clusters.

\subsection{Column Densities and Densities} 
\label{densities}
To determine the molecular abundances relative to molecular hydrogen, we also 
need a measure of the total H$_2$ column density.  We estimate N(H$_2$) from 
the MSX images convolved to match the FCRAO beam resolution and the 
simple relation $\tau_{\lambda}$ = $\sigma_{\lambda}$ $\cdot$ N(H$_{2}$), where $\tau_{\lambda}$ is the dust opacity, $\sigma_{\lambda}$ is the dust extinction cross section, and N(H$_{2}$) is the column density of molecular hydrogen.  
The behavior of the mid-infrared extinction law is an area of active research.  Recent results from Indebetouw et al. (2005), show agreement with Weingartner \& Draine (2001) (R$_v$ = 5.5, ``case B''), we therefore adopt a value for $\sigma_{\lambda}$ at 8.8~$\mu$m of  2.3~$\times$~10$^{-23}$ cm$^{2}$, though this value can be considered reliable only within a factor of 2.
The opacity, $\tau_{\lambda}$, is roughly estimated by examining the relative intensities of the average background ($I_{o,\lambda}$) and central core ($I_{\lambda}$), assuming that $I_{\lambda}$~=~$I_{o,\lambda}e^{-\tau_{\lambda}}$ (i.e. no emission by the core itself).  
The ratio $I_{\lambda}/I_{o,\lambda}$ is related, but not identical, to the brightness contrast 
listed in Table~1, since the values listed in Table~1 do not incorporate the 
convolution to the FCRAO beam, and are provided at the original MSX resolution.

The column densities of C$^{18}$O and CS are estimated by 
assuming that the cores are in local thermodynamic equilibrium (LTE) at a temperature of 15 K, and that the emission is optically thin (Table~4).  While the optically thin assumption is probably 
reasonable for C$^{18}$O, it probably does not hold for CS, which generally has optically thick emission in the interstellar medium.  
For N$_2$H$^+$, the fits to the hyperfine components generally suggest 
low optical depth, or $\tau \sim$1.  However, its emission is likely not in LTE.  At a density of 10$^5$ cm$^{-3}$, the fractional population will be underestimated by a factor of $\sim$1.7 relative to LTE; we therefore apply this correction factor. 
We also crudely estimate the gas density by assuming the cloud is spherical 
and dividing the H$_2$ column density by the diameter of the cloud (using the 
sizes listed in Table~1 and the distances in Table~3).  This gives an 
average density of $\approx 5000$ cm$^{-3}$. This is well below to the
average densities found in other studies of regions of massive star formation
using other tracers.  For example, Plume et al. (1997) surveyed multiple 
transitions of CS in 150 H$_2$O masers (used as signposts of massive star 
formation) and found an average gas density of $7.9 \times 10^5$ cm$^{-3}$.  
An obvious explanation for the lower density in our sample is that the clouds
are not spherical, as we have naively assumed, and may instead be clumpy on scales below our resolution.  Moreover, these objects are likely at an earlier evolutionary state which is characterized by lower densities.  A more detailed analysis of the density and column density will be presented in future work (Gibson et al. 2007, in prep.).  

\subsection{Masses} 
\label{mass}

The total mass of each dark cloud is estimated using an assumed (``near'') distance, an approximate size based on the extent of molecular emission, the molecular column densities and approximate abundance calculated at the peak of absorption.
Table~4 lists the masses of the objects for which there is a significant detection of N$_2$H$^+$ or C$^{18}$O (or both) and a distance could be assigned.
Since some of the dark clouds shown in Figure~1 have a different structure when 
viewed in different molecular tracers and may have different opacities, 
we have, for completeness, estimated the masses independently based on 
the abundances derived from both the N$_{2}$H$^{+}$ and C$^{18}$O data.  Our average mass is 
$\approx 2000 - 3000$ M$_{\odot}$ (depending on whether the N$_2$H$^+$ or 
C$^{18}$O mass is used).

Figure~\ref{fig:mass} shows how the masses of our sample compare to 
the high-mass protostellar objects (HMPOs) presented in Williams et al. (2004),
which were determined from submillimeter continuum emission.  
The masses for the HMPO sample are shown for both their 
``near'' and ``far'' kinematic distances.  Assuming that the sample encompasses objects at both the ``near'' and ``far'' kinematic distances, then the derived masses for HMPOs show a comparable range to our sources.  
%The mass range derived from the N$_{2}$H$^{+}$ emission more closely corresponds to that of Williams et al., which utilizes dust continuum emission, as N$_{2}$H$^{+}$ tends to trace regions of higher gas density that often correspond to peaks of dust continuum emission.  
%The C$^{18}$O emission, however, will predominantly trace a larger but lower density envelope surrounding the dense core.  The larger volume sampled by the C$^{18}$O yields a larger mass, which is reflected in our results.
One caveat with our comparison to the Williams et al. sample lies in the 
selection bias of our sample.  As we will discuss below, our observations are only sensitive to relatively massive objects when the cores reside at such large distances.  Furthermore, the HMPOs in the Williams et al. study also contain protostars which heat the surrounding 
environment and increase the dust emission.  Therefore, in the warmer 
environments of the Williams et al. survey, lower mass cores would be easier 
to detect.  The mass distribution shown in Figure~\ref{fig:mass} also
shows good agreement with that found by Shirley et al. (2003), who observed
CS emission from a sample of massive star forming regions and found a mean 
mass of 920 M$_{\odot}$ with a large dispersion.

Several assumptions contribute to the uncertainty in the mass calculation, which is dominated by the error in the abundance calculation.  We assume a constant temperature of 15~K, and a 5~K change in this value results in a $\sim$20\% change in the abundance.  The uncertainty in the dust opacity/column density relation contributes another factor of 2.
%Additionally, while we were careful to account for the most recent advances in the understanding of the extinction correction, there is still considerable uncertainty in this area, and we estimate that our choice of $\sigma_{\lambda}$ is likely accurate within a factor of $\sim$2.  
Finally, we assume a constant abundance along the line of sight, which likely contributes an additional factor of 2 -- 3 to the mass estimates.  

We note that the typical distance to these clouds is $\sim$4 kpc, and with 50$\arcsec$ resolution, we are likely only sensitive to objects of some minimum mass.  To examine this limit, we modeled the emission of a cloud assuming a constant density of 10$^5$cm$^{-3}$, a radius of 0.1pc, an N$_2$H$^+$ abundance of 5 $\times$ 10$^{-10}$ using a Monte Carlo radiation transfer model (Ashby et al. 2000).  We estimate that our observations are capable of detecting clouds of mass greater than 50-100 M$_{\odot}$ at a distance of 4 kpc with a 50$\arcsec$ beam. 

Assuming that the ``near'' distance assumption is correct, the uncertainties mentioned above can account for up to a factor of $\sim4-6$ in mass error, as the 14\% error in translating the galactic rotation curve to kinematic distances only introduces a distance error of $\sim$20\%.  This accounting suggests that these objects are at least 100~M$_{\odot}$, and likely an order of magnitude more massive.  Should the ``near'' distance assumption be incorrect and the dark clouds lie closer to the ``far'' kinematic distance, then the distance error dominates the calculation, and these clouds are substantially more massive.

%Our sources were chosen because they were the most compact and highest opacity objects near other star formation sites on the plane of the sky, and we would expect to detect massive clouds more frequently than low-mass clouds at these distances.  

\subsection{Velocity Dispersion} 
\label{velocity}
The width of emission lines in star forming clouds serves as a useful 
diagnostic in determining the nature of a molecular region.  According to 
Goldsmith (1987), the sites of massive star formation, GMCs, are characterized 
by large linewidths, while the isolated sites of low-mass star formation,
dark clouds, have considerably smaller linewidths.  We illustrate this range 
in Figure~\ref{fig:vel}.  

Caselli et al. (1998) derived linewidths for N$_2$H$^+$ for a sample of 
low-mass, dense clumps in dark clouds, a site in which we expect to find 
narrow lines.  The average linewidth in the Caselli et al. study was 0.33 km~s$^{-1}$ for clumps in which no IRAS source is detected.  Current chemical models and observations indicate that NH$_3$ 
and N$_2$H$^+$ are related because NH$_3$ likely forms via pathways linked 
N$_2$H$^+$ (Aikawa et al. 2005).  Therefore, we use the Harju et al. 
(1993) linewidths for NH$_3$ clumps in Orion and Cepheus, known to be large 
regions of clustered, high-mass star formation for comparison.  In addition, we include a comparison with linewidths of a sample of ammonia cores presented in Molinari et al. (1996), though we only include only the ``Low'' sources, a sample they argue have less luminous IRAS source, more quiescent envelopes, and, therefore, are younger than their ``High'' counterparts.  In our sample, we find average linewidths of 2.0 km~s$^{-1}$ for N$_2$H$^+$, 2.1 km~s$^{-1}$ for C$^{18}$O, and 2.9 km~s$^{-1}$ for CS.
Though we see a broad range of line widths in the N$_2$H$^+$ observations
of the dark clouds, the characteristic line widths presented here are 
generally higher than those of the Harju et al. study and, to a greater 
extent, the Caselli et al. study, which implies that the objects in this 
sample are likely not associated with low-mass star formation.  However, we find good agreement with the Molinari et al. sample. 

For the CS $J = 2 \rightarrow 1$ transition, the linewidths in the dark cloud
sample are narrower than those observed by Plume et al. (1997) in survey of 
massive star forming regions, which averaged 4.2 km~s$^{-1}$ in this line.
However, since Plume et al. surveyed regions known to have undergone massive 
star formation, it is possible that the current generation of massive stars 
are injecting additional turbulence into the surrounding environment.  
Similarly, the Shirley et al. (2003) study observed the CS $J = 5 \rightarrow 4$ transition in star-forming cores and found linewidths averaging 5.6 km~s$^{-1}$.
The narrower lines in the dark clouds indicates that they are still relatively 
quiescent and suggest that may, indeed, be pre-cursor sites of intermediate or massive star 
formation.

\section{Summary} 
\label{summary}
We have identified 41 infrared dark (at 8.8 $\mu$m) clouds that are opaque, 
compact, and associated with UC HII regions using MSX survey data.
In order to determine some basic characteristics of these dark clouds, we have 
mapped emission from N$_{2}H^{+}$ $1 \rightarrow 0$, CS $2 \rightarrow 1$ and 
C$^{18}$O $1 \rightarrow 0$ using the FCRAO.  
The morphology and relative strengths of these molecular lines varies 
dramatically, possibly indicating evolutionary differences and/or the 
presence of undetected embedded protostar(s).  
Based on the derived kinematic distances and the simplifying 
assumption that the cores are optically thin, we have determined 
average properties: diameter $<D> \approx 0.9$ pc, 
density $<n> \approx 5000$ cm$^{-3}$, and mass 
$<M> \approx 2500$ M$_{\odot}$.  The low density estimate likely 
indicates that the dark clouds are clumpy rather than homogeneous.
The derived masses, however, are comparable to those derived for a sample of HMPOs.  
The linewidths are larger than those seen in low-mass star forming cores 
and larger than in high-mass star forming cores in 
Orion and Cepheus.  However, they are narrower than the 
CS linewidths seen in regions that are actively forming massive stars . These observations, taken together, suggest that the infrared dark clouds may be the relatively quiescent pre-cursors to intermediate or
massive star formation, the so-called ``pre-protostellar cores.''  At present, we do not know if these sources are truly starless.  Only by obtaining very sensitive infrared observations can we confirm the starless nature of these cores.

\acknowledgements
We are grateful to the referee for a very thorough report, which improved this paper.  We are grateful to J. Weingartner for providing detailed dust opacity coefficients.  This research has made use of the SIMBAD database, operated at CDS, Strasbourg, France

\begin{appendix}
%\appendix 
\section{Coincident Sources}

The coordinates of each dark region were examined by the {\it Set of Identifications, 
Measurements, and Bibliography for Astronomical Data (SIMBAD)} database in search of associated objects indicative of active star formation (e.g. masers, IRAS sources, radio sources).  In most cases, the searches yielded no results, but some revealed objects within one arcminute of the dark region's coordinates.  Below, we summarize any associations with the objects for which we present emission maps.  We note that whenever coincident sources are found, we provide the offsets relative to the central absorption peak position given in Table 1.

\medskip

%\noindent \textit{G05.85$-$0.23} No coincident sources known.
%\newline
%\noindent \textit{G06.26$-$0.51} No coincident sources known.
%\newline
%\noindent \textit{G09.16$+$1.16} No coincident sources known.
%\newline
\noindent \textit{G09.21$-$0.22} There is an IRAS source (IRAS 18038-2105; $\alpha (2000) = 18^{h} 06^{m} 53.1^{s}$, $\delta (2000) = -21^{\circ} 04 \arcmin 38 \arcsec$) in the vicinity of this absorbing cloud, offset by $0 \farcm 78$ from the center of the dark region.
\newline
%\noindent \textit{G09.28$-$0.15} No coincident sources known.
%\newline
%\noindent \textit{G09.86$-$0.04} No coincident sources known. 
%\newline
\noindent \textit{G09.88$-$0.11} This region has an associated 1612 MHz OH maser which is cited in Blommaert, Van Langevelde, and Michaels (1994).  This source, OH 9.878-0.127, has a position of   $\alpha (2000) = 18^{h} 07^{m} 59 \fs 07$, $\delta (2000) = -20^{\circ} 27 \arcmin 34 \farcs 3$, which is offset from the center of our region by $1 \farcm 05$.  This object may be associated with a circumstellar shell around an evolved star, and the velocities believed to correspond to the expanding shells of material are 79.5 and 111.3 km~s$^{-1}$, which is not coincident with the velocity of the emission detected here (17 km~s$^{-1}$).  
\newline
\noindent \textit{G10.59$-$0.31} This region contains a radio source, located at $\alpha (2000) = 18^{h} 10^{m} 6 \fs 18$, $\delta (2000) = -19^{\circ} 55 \arcmin 33 \farcs 11$, 
according to Zoonematkermani et al.\ (1990), which is offset from the center of our region
by $0 \farcm 44$.  
%At 1400 MHz, the peak flux density is 41 Jy, the integrated flux density is 42 Jy, and the source diameter is $ 0 \farcs 9$. 
\newline
%\noindent \textit{G10.70$-$0.33} No coincident sources known.
%\newline
%\noindent \textit{G10.99$-$0.09} No coincident sources known.
%\newline
%\noindent \textit{G12.22$+$0.14} No coincident sources known.
%\newline
\noindent \textit{G12.50$-$0.22} There is an IRAS source (IRAS 18197-1812; $\alpha (2000) = 18^{h} 13^{m} 39 \fs 0$, $\delta (2000) = -18^{\circ} 11 \arcmin 46 \arcsec$) offset by $0 \farcm 96$ from the absorbing region.
\newline
%\noindent \textit{G14.33$-$0.57} No coincident sources known.
%\newline
\noindent \textit{G19.37$-$0.03} This region is near a known UC HII region, with a water and methanol maser (Codella \& Felli 1995; Szymczak et al. 2000) also identified in the vicinity ($\alpha (2000) = 18^{h} 26^{m} 24 \fs 3$, $\delta (2000) = -12^{\circ} 3 \arcmin 46 \arcsec$, offset $0 \farcm 85$ from the absorbing region.  The peak velocity of this maser is 26.3 km~s$^{-1}$ (Szymczak et al. 2000), which is consistent with the velocity of our measured emission (27 km~s$^{-1}$).  Molinari et al. (1996) also observed this maser site and designated it as Mol 55.
\newline
%\noindent \textit{G19.40$-$0.01} No coincident sources known.
%\newline
%\noindent \textit{G23.37$-$0.29} No coincident sources known. 
%\newline
%\noindent \textit{G23.48$-$0.53} No coincident sources known.
%\newline
%\noindent \textit{G24.05$-$0.22} No coincident sources known.
%\newline
%\noindent \textit{G24.16$+$0.08} No coincident sources known.
%\newline
%\noindent \textit{G25.99$-$0.06} No coincident sources known.
%\newline
%\noindent \textit{G30.14$-$0.07} No coincident sources known.
%\newline
%\noindent \textit{G30.53$-$0.27} No coincident sources known.
%\newline
\noindent \textit{G30.89$+$0.14} This region has an associated methanol maser, as described by Szymczak et al. (2000).  The maser is located at $\alpha (2000) = 18^{h} 47^{m} 14 \fs 99$, $\delta (2000) = -1 ^{\circ} 44 \arcmin 7 \farcs 99$, which is offset from the center of our region by $0 \farcm 99$.  Szymczak et al. (2000), using a 6.7 GHz survey, measured the internal velocity of the 
maser source to be $\approx 105$ km/s, the velocity of the peak to be 101.5 km/s.  This is consistent with one of the velocity components we measured in this object (108 km~s$^{-1}$).  
%They also measured a peak flux density of 33 Jy, and the integrated flux density of 60.5 Jy. 
\newline
%\noindent \textit{G30.98 $-$0.15} No coincident sources known.
%\newline
%\noindent \textit{G31.02$-$0.12} No coincident sources known.
%\newline
\noindent \textit{G32.01$+$0.05} This region has an associated IRAS point source (18470-0050) which is offset from the center of our region by $1 \farcm 00$ at $\alpha (2000) = 18^{h} 49^{m} 36 \fs 6$, $\delta (2000) = -00 ^{\circ} 46 \arcmin 51 \arcsec$ which coincides with a radio source (Becker et al.\, 1994).  This was confirmed to be a methanol maser (observed in the velocity range between 91 and 102 km~s$^{-1}$) by van der Walt et al. (1995), which is consistent with the emission we observed at 95 km~s$^{-1}$.
\newline
%\noindent \textit{G33.82$-$0.22} No coincident sources known.
%\newline
%\noindent \textit{G34.63$-$1.03} No coincident sources known.
%\newline
\noindent \textit{G34.74$-$0.12}  This region has an associated IRAS point source (18526+0130) which is offset from the center of our region by $0 \farcm 80$ at $\alpha (2000) = 18^{h} 55^{m} 10^{s}$, $\delta (2000) = +1 ^{\circ} 33 \arcmin 57 \arcsec$.
%\newline
%\noindent \textit{G34.78 $-$0.80} No coincident sources known.
%\newline
%\noindent \textit{G35.20$-$0.72} No coincident sources known.
%\newline
%\noindent \textit{G37.44$+$0.14} No coincident sources known.
%\newline
%\noindent \textit{G37.89$-$0.15} No coincident sources known.
%\newline
%\noindent \textit{G50.07$+$0.06} No coincident sources known.

\end{appendix}

%\newpage

\clearpage
\LongTables
%
%    SAO Summer Research
%

%\documentclass[preprint]{aastex}
%\begin{document}
\clearpage
\begin{deluxetable}{llccccccccc}
\label{coreprops}
\tabletypesize{\scriptsize}
\tablecolumns{12}
\tablewidth{0pc}
\tablecaption{Basic Properties of the Target MSX Pre-Stellar Cores}
\tablehead{
\colhead{}    & \colhead{}     & \multicolumn{2}{c}{Galactic} & \colhead{}  & \colhead{} &
\colhead{}    & \colhead{}     & \colhead{}     &\colhead{}    & \colhead{}   \\
\colhead{H II}    & \colhead{} & \multicolumn{2}{c}{Coordinates}     & \colhead{} &  \multicolumn{2}{c}{Position} & 
\colhead{Major}    & \colhead{Minor} & \colhead{Brightness} & \colhead{}   \\
\cline{3-4} \cline{6-7} 
\colhead{Region}    & \colhead{Name}      & \colhead{$l$}           & \colhead{$b$}  & \colhead{}   & \colhead{$\alpha (2000)$} &
\colhead{$\delta (2000)$} &
\colhead{Axis}   & \colhead{Axis} & \colhead{Contrast} & \colhead{Bandwidth\tablenotemark{a}}   \\
\colhead{}        & \colhead{}     & \colhead{$(\degr )$}      & \colhead{$(\degr )$}  & \colhead{}   & \colhead{$(h:m:s)$}&
\colhead{$(\degr:\arcmin:\arcsec)$} & \colhead{$(\arcmin)$} & \colhead{$(\arcmin)$} & \colhead{} & \colhead{(km s$^{-1})$}  }
\startdata
G0589 & G5.85$-$0.23  & 5.854  &  $-$0.231  & &  17:59:49.50 &  $-$24:00:50.25   & 0.64 & 0.60 & 0.409 & $-20\,-\,112$  \\*
      & G6.01$-$0.38  & 6.013  &  $-$0.384  & &  18:00:45.13 &  $-$23:57:07.10   & 1.90 & 0.78 & 0.427 &\nodata  \\*
      & G6.26$-$0.51  & 6.263  &  $-$0.511  & &  18:01:46.57 &  $-$23:47:51.61   & 1.38 & 0.93 & 0.587 & $-25\,-\,107$  \\ 
G0867 & G8.56$+$0.42  & 8.559  & \phs 0.415 & &  18:03:11.72 &  $-$21:20:37.48   & 0.40 & 0.36 & 0.286 & \nodata   \\*
      & G8.64$-$0.09  & 8.636  &  $-$0.086  & &  18:05:13.98 &  $-$21:31:21.58   & 0.60 & 0.35 & 0.292 & \nodata  \\*
      & G9.16$+$0.06  & 9.156  & \phs 0.060 & &  18:05:46.71 &  $-$20:59:51.77   & 0.80 & 0.46 & 0.340 &$-25\,-\,107$  \\*
      & G9.20$-$0.20  & 9.204  &  $-$0.199  & &  18:06:50.86 &  $-$21:04:56.00   & 0.80 & 0.47 & 0.288 & \nodata  \\*
      & G9.21$-$0.22  & 9.213  &  $-$0.217  & &  18:06:56.03 &  $-$21:04:59.32   & 0.80 & 0.60 & 0.313 & $-35\,-\,113$   \\*
      & G9.28$-$0.15  & 9.281  &  $-$0.152  & &  18:06:49.96 &  $-$20:59:31.46   & 1.60 & 0.64 & 0.431 & $-38\,-\,116$   \\*
      & G9.64$+$0.18  & 9.626  & \phs 0.184 & &  18:06:17.86 &  $-$20:31:37.28   & 0.23 & 0.15 & 0.120 & \nodata   \\ 
G1030 & G9.80$-$0.15  & 9.796  &  $-$0.151  & &  18:07:54.07 &  $-$20:32:30.17   & 0.90 & 0.68 & 0.242 & \nodata  \\* 
      & G9.85$-$0.14  & 9.846  &  $-$0.139  & &  18:07:57.60 &  $-$20:29:31.81   & 1.20 & 0.80 & 0.254 & \nodata  \\*
      & G9.86$-$0.04  & 9.856  &  $-$0.039  & &  18:07:36.47 &  $-$20:26:05.34   & 1.24 & 0.61 & 0.371 & $-35\,-\,113$ \\*
      & G9.88$-$0.11  & 9.881  &  $-$0.111  & &  18:07:55.69 &  $-$20:26:52.74   & 1.10 & 0.91 & 0.403 & $-35\,-\,113$ \\*
      & G10.27$+$0.19 & 10.274 & \phs 0.186 & &  18:07:38.17 &  $-$19:57:36.60   & 0.56 & 0.53 & 0.183 & \nodata   \\*
      & G10.59$-$0.31 & 10.588 &  $-$0.311  & &  18:10:08.02 &  $-$19:55:34.98   & 1.00 & 0.48 & 0.295 & $-35\,-\,113$  \\*
      & G10.70$-$0.33 & 10.704 &  $-$0.325  & &  18:10:25.44 &  $-$19:49:53.70   & 1.00 & 0.84 & 0.367 & $-60\,-\,86$  \\*
      & G10.74$-$0.13 & 10.743 &  $-$0.127  & &  18:09:46.03 &  $-$19:42:06.37   & 5.20 & 1.50 & 0.600 & \nodata  \\*
      & G10.74$+$0.01 & 10.743 & \phs 0.005 & &  18:09:16.59 &  $-$19:38:16.41   & 1.23 & 0.93 & 0.413 & \nodata \\*
      & G10.99$-$0.09 & 10.989 &  $-$0.086  & &  18:10:07.18 &  $-$19:27:59.72   & 0.80 & 0.58 & 0.593 & $-60\,-\,86$ \\*
      & G11.13$+$0.11 & 11.126 & \phs 0.114 & &  18:09:39.49 &  $-$19:14:59.92   & 0.58 & 0.45 & 0.239 & \nodata  \\*
      & G11.13$-$0.13 & 11.131 &  $-$0.129  & &  18:10:34.21 &  $-$19:21:46.84   & 2.40 & 0.60 & 0.463 & \nodata  \\*
      & G11.23$+$0.07 & 11.226 & \phs 0.068 & &  18:10:02.01 &  $-$19:11:04.88   & 1.22 & 0.27 & 0.202 & \nodata  \\*
      & G11.24$+$0.07 & 11.241 & \phs 0.066 & &  18:10:04.30 &  $-$19:10:21.11   & 1.03 & 0.52 & 0.283 & \nodata  \\ 
G1194 & G12.22$+$0.14 & 12.219 & \phs 0.138 & &  18:11:47.79 &  $-$18:16:51.43   & 1.68 & 0.80 & 0.388 & $-35\,-\,113$  \\*
      & G12.50$-$0.22 & 12.498 &  $-$0.222  & &  18:13:41.44 &  $-$18:12:32.16   & 0.88 & 0.75 & 0.475 & $-35\,-\,113$  \\*
      & G12.58$+$0.27 & 12.578 & \phs 0.273 & &  18:12:01.49 &  $-$17:54:04.62   & 1.55 & 0.61 & 0.263 & \nodata  \\*
      & G12.61$+$0.25 & 12.608 & \phs 0.248 & &  18:12:10.65 &  $-$17:53:13.15   & 0.86 & 0.59 & 0.235 & \nodata  \\*
      & G12.74$-$0.36 & 12.736 &  $-$0.361  & &  18:14:41.01 &  $-$18:03:59.01   & 0.87 & 0.52 & 0.280 & \nodata  \\ 
G1504 & G14.33$-$0.57 & 14.334 &  $-$0.567  & &  18:18:37.67 &  $-$16:45:30.54   & 0.97 & 0.51 & 0.362 & $-54\,-\,93$  \\*
      & G14.38$-$0.46 & 14.384 &  $-$0.457  & &  18:18:19.32 &  $-$16:39:44.47   & 1.36 & 0.76 & 0.268 & \nodata   \\*
      & G14.40$-$0.60 & 14.398 &  $-$0.604  & &  18:18:53.43 &  $-$16:43:10.49   & 1.51 & 1.51 & 0.412 & \nodata  \\*
      & G14.49$-$0.15 & 14.493 &  $-$0.146  & &  18:17:23.71 &  $-$16:25:08.20   & 1.68 & 0.90 & 0.465 & \nodata  \\*
      & G14.64$-$0.57 & 14.636 &  $-$0.571  & &  18:19:14.32 &  $-$16:29:39.47   & 2.25 & 0.90 & 0.568 & \nodata  \\*
      & G14.65$-$0.18 & 14.648 &  $-$0.117  & &  18:17:35.72 &  $-$16:16:07.42   & 0.70 & 0.32 & 0.346 & \nodata  \\*
      & G15.05$+$0.09 & 15.053 & \phs 0.089 & &  18:17:38.41 &  $-$15:48:52.11   & 1.60 & 0.63 & 0.532 & \nodata  \\*
      & G15.55$-$0.45 & 15.548 &  $-$0.447  & &  18:20:34.52 &  $-$15:37:54.30   & 0.88 & 0.48 & 0.280 & \nodata  \\ 
G1961 & G19.25$-$0.07 & 19.246 &  $-$0.069  & &  18:26:20.57 &  $-$12:11:16.74   & 0.70 & 0.39 & 0.238 & \nodata  \\*
      & G19.28$-$0.39 & 19.281 &  $-$0.387  & &  18:27:33.68 &  $-$12:18:18.35   & 1.08 & 0.78 & 0.417 & $-35\,-\,113$  \\*
      & G19.29$+$0.08 & 19.291 & \phs 0.078 & &  18:25:53.80 &  $-$12:04:46.57   & 2.75 & 1.10 & 0.447 & \nodata  \\*
      & G19.35$-$0.04 & 19.348 &  $-$0.039  & &  18:26:25.72 &  $-$12:05:01.46   & 0.72 & 0.68 & 0.275 & \nodata \\*
      & G19.37$-$0.03 & 19.374 &  $-$0.034  & &  18:26:27.61 &  $-$12:03:30.32   & 1.01 & 0.80 & 0.375 & $-35\,-\,113$   \\*
      & G19.40$-$0.01 & 19.396 &  $-$0.006  & &  18:26:24.05 &  $-$12:01:33.28   & 1.05 & 0.94 & 0.330 & $-35\,-\,113$  \\*
      & G19.91$-$0.21 & 19.913 &  $-$0.207  & &  18:28:06.69 &  $-$11:39:42.83   & 1.39 & 0.98 & 0.346 & \nodata  \\*
      & G19.97$-$0.11 & 19.967 &  $-$0.106  & &  18:27:50.94 &  $-$11:34:01.67   & 0.65 & 0.65 & 0.275 & \nodata  \\*
      & G19.98$-$0.21 & 19.979 &  $-$0.212  & &  18:28:15.29 &  $-$11:36:20.91   & 1.57 & 0.72 & 0.318 & \nodata  \\ 
G2371 & G23.32$+$0.06 & 23.318 & \phs 0.056 & &  18:33:34.18 &  $-$08:31:19.32   & 2.10 & 0.42 & 0.259 & \nodata  \\*
      & G23.37$-$0.29 & 23.368 &  $-$0.287  & &  18:34:53.66 &  $-$08:38:08.59   & 0.72 & 0.50 & 0.376 & \phs $37\,-\,183$  \\*
      & G23.38$-$0.13 & 23.378 &  $-$0.126  & &  18:34:20.09 &  $-$08:33:09.71   & 0.77 & 0.40 & 0.321 & \nodata  \\*
      & G23.38$-$0.51 & 23.381 &  $-$0.521  & &  18:35:45.56 &  $-$08:43:54.76   & 0.71 & 0.60 & 0.227 & \nodata  \\*
      & G23.38$+$0.29 & 23.383 & \phs 0.288 & &  18:32:51.52 &  $-$08:21:26.29  & 0.57 & 0.50 & 0.311 & \nodata   \\*
      & G23.44$-$0.52 & 23.441 &  $-$0.524  & &  18:35:52.89 &  $-$08:40:47.92   & 1.88 & 0.53 &0.313 & \nodata  \\*
      & G23.48$+$0.11 & 23.476 & \phs 0.106 & &  18:33:41.10 &  $-$08:21:31.62   & 1.22 & 0.54 & 0.378 & \nodata  \\*
      & G23.48$-$0.53 & 23.481 &  $-$0.534  & &  18:35:59.52 &  $-$08:38:56.65   & 1.50 & 0.50 & 0.480 & $-35\,-\,113$ \\*
      & G23.61$-$0.01 & 23.614 &  $-$0.009  & &  18:34:21.28 &  $-$08:17:21.61   & 1.82 & 0.76 & 0.471 & \nodata  \\*
      & G24.05$-$0.22 & 24.051 &  $-$0.216  & &  18:35:54.61 &  $-$07:59:48.07   & 0.81 & 0.51 & 0.368 & \phs $37\,-\,184$ \\*
      & G24.16$+$0.08 & 24.158 & \phs 0.078 & &  18:35:03.29 &  $-$07:45:58.96   & 0.90 & 0.84 & 0.390 & \phs $37\,-\,184$  \\*
      & G24.17$+$0.06 & 24.170 & \phs 0.064 & &  18:35:07.65 &  $-$07:45:43.85   & 0.71 & 0.37 & 0.344 & \nodata  \\*
      & G24.37$-$0.16 & 24.369 &  $-$0.157  & &  18:36:17.34 &  $-$07:41:13.82   & 1.68 & 1.56 & 0.467 & \nodata  \\*
      & G24.37$-$0.21 & 24.371 &  $-$0.214  & &  18:36:29.82 &  $-$07:42:41.72   & 2.10 & 0.96 & 0.258 & \nodata  \\*
      & G24.44$-$0.23 & 24.436 &  $-$0.234  & &  18:36:41.36 &  $-$07:39:46.99   & 1.54 & 1.51 & 0.546 & \nodata  \\*
      & G24.63$+$0.15 & 24.629 & \phs 0.148 & &  18:35:40.73 &  $-$07:18:57.64   & 2.15 & 0.70 & 0.533 & \nodata  \\ 
G2572 & G25.08$+$0.20 & 25.076 & \phs 0.199 & &  18:36:19.50 &  $-$06:53:44.33   & 0.80 & 0.63 & 0.363 & \nodata  \\*
      & G25.24$-$0.22 & 25.243 &  $-$0.216  & &  18:38:07.16 &  $-$06:56:16.48   & 0.72 & 0.56 & 0.298 & \nodata  \\*
      & G25.25$-$0.24 & 25.248 &  $-$0.236  & &  18:38:12.01 &  $-$06:56:33.53   & 1.00 & 0.50 & 0.283 & \nodata  \\*
      & G25.61$+$0.24 & 25.613 & \phs 0.239 & &  18:37:10.53 &  $-$06:24:01.14   & 1.70 & 1.70 & 0.374 & \nodata  \\*
      & G25.99$-$0.06 & 25.985 &  $-$0.057  & &  18:38:55.26 &  $-$06:12:20.39   & 1.03 & 0.80 & 0.337 & $-35\,-\,113$  \\ 
G2996 & G30.14$-$0.07 & 30.135 &  $-$0.069  & &  18:46:34.71 &  $-$02:31:13.90   & 1.40 & 0.60 & 0.287 &$-35\,-\,113$  \\*
      & G30.31$-$0.28 & 30.309 &  $-$0.282  & &  18:47:39.31 &  $-$02:27:46.38   & 0.82 & 0.63 & 0.226 & \nodata  \\*
      & G30.49$-$0.39 & 30.494 &  $-$0.392  & &  18:48:23.10 &  $-$02:20:54.35   & 0.84 & 0.81 & 0.414 & \phs $22\,-\,168$  \\*
      & G30.53$-$0.27 & 30.353 &  $-$0.272  & &  18:47:42.00 &  $-$02:25:08.99   & 1.25 & 0.91 & 0.310 & $-35\,-\,113$   \\*
      & G30.58$-$0.25 & 30.579 &  $-$0.252  & &  18:48:02.48 &  $-$02:12:32.12   & 1.53 & 0.58 & 0.305 & \nodata \\*
      & G30.66$+$0.05 & 30.663 & \phs 0.046 & &  18:47:08.00 &  $-$01:59:53.60   & 1.82 & 0.80 & 0.358 & \nodata  \\*
      & G30.69$+$0.06 & 30.689 & \phs 0.056 & &  18:47:08.72 &  $-$01:58:13.85   & 0.75 & 0.63 & 0.217 & \nodata  \\ 
G3141 & G30.89$+$0.14 & 30.894 & \phs 0.136 & &  18:47:14.08 &  $-$01:45:05.75   & 1.60 & 0.60 & 0.321 & $-35\,-\,113$  \\*
      & G30.98$-$0.15 & 30.978 &  $-$0.148  & &  18:48:23.97 &  $-$01:48:23.07   & 1.33 & 0.80 & 0.411 & $-35\,-\,113$ \\*
      & G31.02$-$0.12 & 31.024 &  $-$0.116  & &  18:48:22.17 &  $-$01:45:03.19   & 1.62 & 0.70 & 0.450 & $-35\,-\,113$  \\*
      & G31.23$+$0.02 & 31.226 & \phs 0.024 & &  18:50:03.85 &  $-$00:37:02.40   & 1.20 & 0.80 & 0.361 & \nodata  \\*
      & G31.39$+$0.30 & 31.391 & \phs 0.296 & &  18:47:34.34 &  $-$01:14:10.89   & 2.10 & 1.70 & 0.349 & \nodata  \\*
      & G31.70$-$0.50 & 31.699 &  $-$0.496  & &  18:50:57.27 &  $-$01:19:24.42   & 0.68 & 0.50 & 0.348 & \nodata  \\*
      & G31.71$-$0.49 & 31.723 &  $-$0.486  & &  18:50:57.76 &  $-$01:17:51.09   & 0.63 & 0.60 & 0.336 & \nodata  \\*
      & G32.01$+$0.05 & 32.013 & \phs 0.056 & &  18:49:33.70 &  $-$00:47:32.25   & 1.17 & 1.10 & 0.526 & $-35\,-\,113$  \\ 
G3350 & G32.84$-$0.03 & 32.843 &  $-$0.032  & &  18:51:23.35 &  $-$00:05:37.42   & 1.30 & 0.97 & 0.357 & \nodata  \\*
      & G33.36$-$0.01 & 33.363 &  $-$0.007  & &  18:52:14.91 & \phs 00:22:49.67  & 1.58 & 0.84 & 0.288 & \nodata  \\*
      & G33.42$+$0.13 & 33.418 & \phs 0.126 & &  18:51:52.52 & \phs 00:29:24.18  & 0.44 & 0.25 & 0.217 & \nodata  \\*
      & G33.70$-$0.02 & 33.699 &  $-$0.016  & &  18:52:53.60 & \phs 00:40:31.43  & 1.80 & 0.50 & 0.250 & \nodata  \\ 
G3426 & G33.82$-$0.22 & 33.819 &  $-$0.219  & &  18:53:50.10 & \phs 00:41:22.54  & 0.90 & 0.53 & 0.432 & \phs $2\,-\,149$  \\*
      & G34.13$+$0.08 & 34.134 & \phs 0.076 & &  18:53:21.57 & \phs 01:06:16.14  & 1.68 & 0.50 & 0.254 & \nodata  \\*
      & G34.26$+$0.19 & 34.263 & \phs 0.189 & &  18:53:11.55 & \phs 01:16:14.93  & 1.75 & 0.61 & 0.271& \nodata  \\*
      & G34.74$-$0.12 & 34.739 &  $-$0.122  & &  18:55:10.11 & \phs 01:33:09.25  & 0.90 & 0.80 & 0.393 & $-15\,-\,131$  \\*
      & G34.74$+$0.01 & 34.744 & \phs 0.006 & &  18:54:43.32 & \phs 01:36:55.46  & 0.75 & 0.57 & 0.212 & \nodata  \\*
      & G35.04$-$0.47 & 35.043 &  $-$0.474  & &  18:56:58.62 & \phs 01:39:44.74  & 0.65 & 0.41 & 0.312 & \nodata  \\ 
G3520 & G34.63$-$1.03 & 34.627 &  $-$1.026  & &  18:58:10.95 & \phs 01:02:25.33  & 0.26 & 0.18 & 0.447 & $-29\,-\,117$  \\*
      & G34.78$-$0.80 & 34.778 &  $-$0.804  & &  18:57:40.08 & \phs 01:16:33.76  & 0.56 & 0.45 & 0.341 & $-35\,-\,114$  \\*
      & G35.02$-$1.50 & 35.018 &  $-$1.497  & &  19:00:34.38 & \phs 01:10:23.16  & 0.97 & 0.35 & 0.375 & \nodata  \\*
      & G35.20$-$0.72 & 35.203 &  $-$0.721  & &  18:58:08.92 & \phs 01:41:31.31  & 0.78 & 0.51 & 0.436 & $-29\,-\,117$ \\ 
G3755 & G37.08$-$0.15 & 37.081 &  $-$0.149  & &  18:59:32.80 & \phs 03:37:25.91  & 0.42 & 0.40 & 0.321& \nodata   \\*
      & G37.25$+$0.01 & 37.253 & \phs 0.011 & &  18:59:17.48 & \phs 03:50:59.94  & 0.98 & 0.52 & 0.296 & \nodata   \\*
      & G37.42$+$0.17 & 37.418 & \phs 0.173 & &  18:59:00.95 & \phs 04:04:14.78  & 0.70 & 0.64 & 0.342 & \nodata   \\*
      & G37.44$+$0.14 & 37.439 & \phs 0.138 & &  18:59:10.75 & \phs 04:04:24.37  & 0.79 & 0.40 & 0.322 & $-35\,-\,113$  \\*
      & G37.89$-$0.15 & 37.886 &  $-$0.152  & &  19:01:02.00 & \phs 04:20:18.23  & 1.03 & 0.58 & 0.325 & $-26\,-\,107$  \\ 
G4318 & G43.19$-$0.16 & 43.187 &  $-$0.162  & &  19:10:53.01 & \phs 09:02:30.58  & 0.42 & 0.20 & 0.304 & \nodata  \\*
      & G43.32$-$0.20 & 43.318 &  $-$0.204  & &  19:11:16.78 & \phs 09:08:18.83  & 0.50 & 0.42 & 0.165 & \nodata  \\*
      & G43.78$+$0.05 & 43.776 & \phs 0.046 & &  19:11:14.38 & \phs 09:39:36.89  & 0.92 & 0.91 & 0.377 & $-23\,-\,124$  \\ 
G4389 & G43.64$-$0.82 & 43.644 &  $-$0.824  & &  19:14:07.05 & \phs 09:08:24.64  & 0.60 & 0.21 & 0.577 & $-19\,-\,127$  \\ 
G4426 & G44.29$-$0.09 & 44.291 &  $-$0.092  & &  19:12:42.25 & \phs 10:03:10.40  & 0.71 & 0.46 & 0.219 & \nodata  \\ 
G5023 & G48.84$+$0.15 & 48.836 & \phs 0.151 & &  19:20:29.90 & \phs 14:11:12.12  & 1.22 & 0.58 & 0.317 & \nodata \\*
      & G48.84$+$0.14 & 48.848 & \phs 0.136 & &  19:20:34.57 & \phs 14:11:24.84  & 0.70 & 0.43 & 0.242 & \nodata  \\*
      & G50.07$+$0.06 & 50.071 & \phs 0.059 & &  19:23:14.36 & \phs 15:13:58.11  & 0.86 & 0.33 & 0.378 & $-13\,-\,133$  \\ 
G5031 & G51.00$-$0.18 & 51.001 &  $-$0.177  & &  19:22:55.92 & \phs 15:56:24.24  & 0.60 & 0.35 & 0.290 & \nodata  \\ 
G5410 & G53.88$-$0.18 & 53.879 &  $-$0.181  & &  19:31:42.73 & \phs 18:27:55.78  & 1.12 & 0.42 & 0.417 & $-30\,-\,116$  \\ 
G6148 & G61.52$+$0.02 & 61.519 & \phs 0.024 & &  19:47:09.72 & \phs 25:13:00.46  & 0.60 & 0.45 & 0.334 & \nodata  \\ 
G7578 & G75.75$+$0.75 & 75.753 & \phs 0.749 & &  20:19:57.75 & \phs 37:39:01.90  & 1.10 & 0.90 & 0.474 & $-33\,-\,113$  \\*
      & G76.38$+$0.63 & 76.381 & \phs 0.626 & &  20:22:17.04 & \phs 38:05:50.23  & 0.40 & 0.30 & 0.318 & $-33\,-\,113$  \\ 
\enddata
\tablenotetext{a}{For observed regions, we give the total velocity range to which the observations were sensitive.  A given object was probed in the same velocity range for each observed transition.}

\end{deluxetable}

%\end{document}
 
\clearpage
\LongTables
\begin{landscape}
%\documentclass[12pt,preprint]{aastex}
%\begin{document}
%\usepackage{lscape}
%\setcounter{page}{}
%\begin{landscape}
\begin{center}
\begin{deluxetable}{lrrrrrrrrrrrrr}
\label{molint}
\tabletypesize{\scriptsize}
%\rotate
\tablecolumns{14}
\tablewidth{8.2in}
\tablecaption{Molecular Line Observations}
\tablehead{
\colhead{} & \colhead{} &\colhead{} &
\multicolumn{3}{c}{N$_2$H$^{+}$ J=1--0} &
\colhead{} &
\multicolumn{3}{c}{C$^{18}$O J=1--0} &
\colhead{} &
\multicolumn{3}{c}{CS J=2--1}  \\
\cline{4-6} \cline{8-10} \cline{12-14} \\ 
\colhead{Source} &
\colhead{$\Delta\alpha$} &
\colhead{$\Delta\delta$} &
\colhead{$\int T_A^*dv$} &
\colhead{V} &
\colhead{$\Delta$v} &
\colhead{} &
\colhead{$\int T_A^*dv$} &
\colhead{V} &
\colhead{$\Delta$v} &
\colhead{} &
\colhead{$\int T_A^*dv$} &
\colhead{V} &
\colhead{$\Delta$v}  \\
\colhead{} &
\colhead{$(')$} &
\colhead{$(')$} &
\colhead{(K km s$^{-1}$)} &
\colhead{(km s$^{-1}$)} &
\colhead{(km s$^{-1}$)} &
\colhead{} &
\colhead{(K km s$^{-1}$)} &
\colhead{(km s$^{-1}$)} &
\colhead{(km s$^{-1}$)} &
\colhead{} &
\colhead{(K km s$^{-1}$)} &
\colhead{(km s$^{-1}$)} &
\colhead{(km s$^{-1}$)}  
}
\startdata
G05.85$-$0.23 &0.4&$-$0.4 &1.21(0.15) &17.2(0.1) &0.8(0.1) &&1.44(0.17) &17.0(0.1) &1.7(0.2) &&0.47(0.10) &16.9(0.2) &2.2(0.3) \\ 
%G05.85$-$0.23b &0.4&$-$0.4&($<$0.13)&\nodata &\nodata &&1.10(0.15)&8.4(0.1)&2.0(0.3)&&($<$0.11)&\nodata&\nodata \\
G06.26$-$0.51 &0.0 &0.0 &($<$0.15)&\nodata &\nodata &&1.69(0.11)&22.7(0.1)&2.3(0.2) &&1.64(0.15)&23.2(0.2)&4.2(0.5)\\
%G06.26$-$0.51b &1.7 &0.8 &($<$0.15)&\nodata &\nodata &&1.42(0.08)&16.7(0.1)&1.4(0.1)&&($<$0.12)&\nodata&\nodata \\
G09.16$+$0.06 &0.0 &0.0 &($<$0.12)&\nodata &\nodata &&1.39(0.09)&31.3(0.1)&1.7(0.1) &&0.40(0.07)&31.3(0.1)&1.2(0.3)\\
G09.21$-$0.22 &0.0 &0.0 &3.59(0.15) &42.8(0.1) &1.8(0.2) &&\nodata &\nodata &\nodata &&1.29(0.11) &42.7(0.1) &2.8(0.3)\\
G09.28$-$0.15 &0.0 &0.0 &3.66(0.10)&41.4(0.1)&1.9(0.4) &&\nodata &\nodata &\nodata &&1.27(0.12)&41.3(0.1)&2.6(0.3)\\
G09.86$-$0.04 &0.0 &0.0 &0.87(0.11)&18.1(0.1)&1.1(0.2) &&\nodata &\nodata &\nodata &&1.52(0.14)&17.8(0.1)&2.3(0.2)\\
G09.88$-$0.11 &0.0 &0.0 &($<$0.13) &\nodata &\nodata &&\nodata &\nodata &\nodata &&0.54(0.15)&17.3(0.2)&1.8(0.6)\\
G10.59$-$0.31 &0.0 &0.0 &($<$0.40) &\nodata &\nodata &&\nodata &\nodata &\nodata &&($<$0.24)&\nodata &\nodata \\
G10.70$-$0.33 &0.0 &0.0 &($<$0.13) &\nodata &\nodata &&\nodata &\nodata &\nodata &&($<$0.13)&\nodata &\nodata \\
G10.99$-$0.09 &0.0 &0.0 &4.24(0.11)&29.6(0.1)&2.4(0.2) &&2.25(0.17)&29.5(0.1)&2.2(0.2) &&1.00(0.12)&29.2(0.3)&4.2(0.5)\\
G12.22$+$0.14 &0.0 &0.0 &3.60(0.08)&39.6(0.1)&1.7(0.1) &&\nodata &\nodata &\nodata &&1.78(0.07)&36.7(0.1)&2.2(0.1) \\
G12.50$-$0.22 &0.0 &0.0 &2.97(0.10)&35.8(0.1)&1.8(0.1) &&1.60(0.13)&35.7(0.1)&1.8(0.2) &&1.40(0.10)&35.6(0.1)&2.0(0.2)\\
G14.33$-$0.57a &0.0 &0.0 &($<$0.14)&\nodata &\nodata &&2.19(0.15)&19.3(0.1)&2.1(0.2)&&1.25(0.13)&19.6(0.1)&2.4(0.3)\\
G14.33$-$0.57b &$-1.7$ &0.8 &1.42(0.14)&20.0(0.1) &1.1(0.2) &&1.81(0.12)&19.9(0.1)&1.6(0.2)&&0.84(0.10)&20.3(0.1)&1.4(0.2)\\
G19.28$-$0.39 &0.0 &0.0 &($<$0.17) &\nodata &\nodata &&\nodata &\nodata &\nodata &&0.26(0.07) &54.0(0.1) &1.1(0.2)\\
G19.37$-$0.03 &0.0 &0.0 &3.62(0.11) &27.3(0.1) &2.5(0.1) &&\nodata &\nodata &\nodata &&3.02(0.07) &27.0(0.1) &3.8(0.1)\\
G19.40$-$0.01 &0.0 &0.0 &0.95(0.11) &27.0(0.1) &1.2(0.3) &&\nodata &\nodata &\nodata &&0.81(0.06) &26.5(0.1) &2.9(0.3)\\
G23.37$-$0.29 &0.0 &0.0 &2.88(0.27)&78.5(0.1)&2.0(0.2)&&3.97(0.21)&78.1(0.1)&2.7(0.2) &&2.30(0.12)&77.8(0.1)&4.8(0.3)\\
%G23.37$-$0.29b & $-$0.8 &$-$2.5&2.17(0.15)&102.5(0.1)&2.4(0.2)&&1.78(0.15)&102.4(0.1)&2.4(0.2)&&3.52(0.12)&102.2(0.1)&2.1(0.1) \\
%G23.37$-$0.29c &$-$0.4&$-$2.9&($<$0.50)&\nodata&\nodata&&($<$0.31)&61.3(0.8)\tablenotemark{a}&\nodata&&($<$0.32)&66.3(0.8)\tablenotemark{a}&\nodata \\
G23.48$-$0.53a &0.0 &0.0 &1.45(0.12)&64.8(0.1)&2.5(0.4) &&\nodata &\nodata &\nodata &&0.90(0.09)&63.9(0.3)&4.8(0.6)\\
G23.48$-$0.53b &-2.1 &-2.1 &1.24(0.12)&62.8(0.1)&2.9(0.4) &&\nodata &\nodata &\nodata &&0.73(0.08)&62.7(0.3)&2.9(0.4)\\
%G23.48$-$0.53c &0.0 &0.0 & ($<$0.10)&\nodata&\nodata&&\nodata&\nodata&\nodata &&0.40(0.11)&75.1(0.9)&7.1(2.2) \\ 
G24.05$-$0.22 &0.0 &0.0 &2.71(0.21)&81.4(0.1)&1.9(0.3)&&2.30(0.10)&81.5(0.1)&2.0(0.1)&&1.00(0.13)&82.0(0.2)&2.8(0.5)\\ 
G24.16$+$0.08 &0.0 &0.0 &($<$0.10)&\nodata &\nodata &&1.64(0.16)&51.8(0.1)&1.9(0.2)&&($<$0.10)&\nodata &\nodata \\
%G24.16$+$0.08b &$-$1.7&2.5&0.94(0.08)&113.7(0.1)&3.0(0.3)&&0.66(0.23)&113.7(0.8)&5.7(3.0)&&2.15(0.25)&114.6(0.3)&4.2(0.6)\\
G25.99$-$0.06 &0.0 &0.0 &0.80(0.15)&89.9(0.3)&1.6(0.5)&&\nodata &\nodata &\nodata &&1.05(0.11)&90.2(0.1) &2.4(0.3) \\
G30.14$-$0.07 &0.0 &0.0 &($<$0.12)&\nodata &\nodata &&\nodata &\nodata &\nodata &&($<$0.15)&86.8(0.2)\tablenotemark{a}
&2.7(0.5)\tablenotemark{a} \\
G30.49$-$0.39 &1.2 &$-$0.8 &($<$0.14) &\nodata &\nodata &&\nodata &\nodata &\nodata &&0.62(0.12) &106.4(0.3) &3.0(0.9)\\
G30.53$-$0.27 &0.0 &0.0 &($<$0.12)&\nodata &\nodata &&\nodata &\nodata &\nodata &&1.73(0.19)&102.9(0.4)&7.3(0.9) \\
G30.89$+$0.14 &0.0 &0.0 &1.29(0.13)&96.5(0.2)&3.4(0.3) &&\nodata &\nodata &\nodata &&0.56(0.10)&95.9(0.3)&3.0(0.5)\\
%G30.89$+$0.14b &0.4 & $-$1.7&0.81(0.28)&(0.1)&2.6(0.4)&&\nodata&\nodata&\nodata&&2.93(0.13)&39.7(0.1)&3.6(0.2)\\
%G30.89$+$0.14c & $-$1.3&0.8&2.28(0.26)&105.6(0.1)&3.2(0.1)&&\nodata&\nodata&\nodata&&2.58(0.14)&105.7(0.1)&5.0(0.3)\\
G30.98$-$0.15 &-0.4 &0.0 &4.56(0.12)&77.9(0.1)&2.6(0.1)&&\nodata &\nodata &\nodata &&2.27(0.11)&77.9(0.1)&4.3(0.2)\\
G31.02$-$0.12 &0.0  &0.0 &1.57(0.07)&76.6(0.1)&2.2(0.2)&&1.24(0.11)&76.2(0.1)&3.0(0.3)&&0.66(0.08)&76.6(0.2)&3.3(0.4)\\
%G31.02$-$0.12b &$-$2.5 &$-$0.4&1.34(0.27)&81.2(0.1)&2.3(0.2)&&3.85(0.17)&80.2(0.1)&4.5(0.2)&&1.90(0.10)&81.3(0.1)&3.4(0.3)\\
%G31.02$-$0.12c &0.8& $-$1.7&0.48(0.12)&90.7(0.1)&1.0(0.3)&&1.30(0.13)&91.2(0.1)&2.3(0.3)&&0.61(0.08)&91.3(0.1)&2.4(0.4)\\
G32.01$+$0.05 &0.0 &0.0 &7.18(0.10) &95.3(0.1) &3.9(0.1) &&4.42(0.12) &97.2(0.1) &4.7(0.2) &&4.03(0.09) &96.0(0.1) &6.8(0.2)\\
G33.82$-$0.22 &0.0 &0.0 &1.03(0.11)&11.3(0.1)&1.0(0.3)&&\nodata &\nodata &\nodata &&0.48(0.09)&11.5(0.1)&1.0(0.2)\\
G34.63$-$1.03 &0.0 &0.0 &1.16(0.15)&13.6(0.2)&2.6(0.6)&&0.58(0.08)&12.8(0.1)&1.1(0.2)&&($<$0.15)&\nodata &\nodata \\
G34.74$-$0.12 &0.0 &0.0 &1.92(0.17)&79.1(0.1)&2.7(0.3)&&3.51(0.14)&78.9(0.2)&2.1(0.1)&&1.11(0.10)&78.9(0.2)&3.8(0.4)\\
G34.78$-$0.80 &0.0 &0.0 &0.33(0.11)&43.2(0.4)&3.0(0.7)&&2.25(0.10)&44.1(0.1)&3.3(0.1)&&1.83(0.07)&43.5(0.1)&3.0(0.2)\\
%G34.78$-$0.80b &3.3&3.3 &($<$0.18)&\nodata&\nodata&&0.93(0.28)&36.9(0.2)&2.7(0.6)&&($<$0.08)&\nodata&\nodata \\
G35.20$-$0.72 &0.0 &0.0 &3.02(0.20) &33.1(0.1) &2.5(0.3) &&2.04(0.12) &33.2(0.1) &1.9(0.1) &&1.91(0.13)&33.2(0.1)&3.4(0.3)\\
G37.44$+$0.14a &0.0 &0.0 &($<$0.10) &\nodata  &\nodata &&1.14(0.06)&40.0(0.1)&1.6(0.1)&&0.77(0.04)&40.1(0.1)&1.4(0.1) \\
G37.44$+$0.14b &$-$2.5 &$-$0.4 &0.65(0.08)&17.8(0.1)&0.5(0.1)&&0.52(0.04)&17.8(0.1)&0.7(0.1)&&0.25(0.03)&17.7(0.1)&0.9(0.1) \\
%G37.44$+$0.14c &$-$1.3&2.5&($<$0.10)&\nodata&\nodata&&1.47(0.15)&85.4(0.3)&5.7(0.7)&&($<$0.09)&\nodata&\nodata \\
G37.89$-$0.15 &0.0 &0.0 &0.45(0.08)&12.9(0.1)&0.7(0.1)&&0.63(0.04)&12.9(0.1)&0.7(0.1)&&0.36(0.04)&13.0(0.1)&0.7(0.1)\\
%G37.89$-$0.15b &$-$0.8&1.7&($<$0.08)&\nodata&\nodata&&0.80(0.11)&64.5(0.3)&4.6(0.7)&&($<$0.09)&\nodata&\nodata\\
G43.64$-$0.82 &0.4 &2.4 &($<$0.25) &\nodata &\nodata &&\nodata &\nodata &\nodata &&0.25(0.05) &85.4(0.1) &0.5(0.1)\\
G43.78$+$0.05 &0.0 &0.0 &($<$0.17) &\nodata &\nodata &&\nodata &\nodata &\nodata &&($<$0.09) &\nodata &\nodata\\
G50.07$+$0.06 &$-$0.8 &$-$0.8 &($<$0.19) &\nodata &\nodata &&\nodata &\nodata &\nodata &&0.73(0.08) &54.8(0.1) &1.5(0.2)\\
G53.88$-$0.18 &0.0 &0.0 &($<$0.21) &\nodata &\nodata &&\nodata &\nodata &\nodata &&($<$0.12) &\nodata &\nodata\\
G75.75$+$0.75 &0.0 &0.0 &($<$0.12) &\nodata &\nodata &&\nodata &\nodata &\nodata &&($<$0.07) &\nodata &\nodata \\
G76.38$+$0.63 &0.0 &0.0 &($<$0.15) &\nodata &\nodata &&\nodata &\nodata &\nodata &&($<$0.07) &\nodata &\nodata \\
\enddata
%\tablenotetext{a}{There exists a very broad feature near $\sim$65km s$^{-1}$, but linewidth could not be calculated due to blending.}
\tablenotetext{a}{Non-detection of CS in single scan.  Average of 25 scans detects a weak line at the
5$\sigma$ level.  The velocity and line width from this average are provided in the table.}
\end{deluxetable}
\end{center}
%\end{landscape}

%\newpage

%\end{document}

\clearpage
\end{landscape}
\clearpage
%\documentclass[12pt,preprint]{aastex}
%\begin{document}

\begin{center}
\begin{deluxetable}{lcllcc}
\label{alldist}
\tabletypesize{\scriptsize}
\tablecolumns{6}
\tablewidth{4.7in}
\tablecaption{Kinematic distances to separated velocity components}
\tablehead{
\colhead{}&
\colhead{Velocity}&
\colhead{Near}&
\colhead{Far}&
\colhead{Adopted}&
\colhead{UCHII region}\\
\colhead{Source}&
\colhead{Component}&
\colhead{Distance}&
\colhead{Distance}&
\colhead{Distance}&
\colhead{Distance}\\
\colhead{} &
\colhead{(km s$^{-1}$)} &
\colhead{(kpc)} &
\colhead{(kpc)} &
\colhead{(kpc)} &
\colhead{(kpc)} 
}
\startdata
G05.85$-$0.23 &17 & 3.14$^{+0.66}_{-0.76}$ & 13.78$^{+0.75}_{-0.67}$ & 3.14 & 2.6 \\
			& 9  & 1.53$^{+0.86}_{-0.96}$ & 15.38$^{+0.97}_{-0.86}$ & & \\
G06.26$-$0.51 & 23 & 3.78$^{+0.59}_{-0.67}$ & 13.12$^{+0.67}_{-0.60}$ & 3.78 & 2.6 \\
			& 17 & 3.01$^{+0.68}_{-0.77}$ & 13.89$^{+0.77}_{-0.68}$ & & \\
G09.16$+$0.06 & 31 & 3.81$^{+0.61}_{-0.69}$ & 12.97$^{+0.69}_{-0.61}$ & 3.81 & 6.2 \\
G09.21$-$0.22 & 43 & 4.57$^{+0.53}_{-0.59}$ & 12.21$^{+0.59}_{-0.53}$ & 4.57 & 6.2 \\
G09.28$-$0.15 & 42 & 4.48$^{+0.54}_{-0.61}$ & 12.30$^{+0.61}_{-0.54}$ & 4.48 & 6.2 \\
G09.86$-$0.04 & 18 & 2.36$^{+0.78}_{-0.88}$ & 14.39$^{+0.87}_{-0.78}$ & 2.36 & 6.0 \\
%G09.88$-$0.11 & 17 & 2.28$^{+0.78}_{-0.89}$ & 14.47$^{+0.89}_{-0.78}$ & 2.28 & 6.0 \\
G10.99$-$0.09 & 30 & 3.31$^{+0.69}_{-0.76}$ & 13.37$^{+0.77}_{-0.68}$ & 3.31 & 6.0 \\
G12.22$+$0.14 & 40 & 3.75$^{+0.65}_{-0.72}$ & 12.86$^{+0.73}_{-0.65}$ & 3.75 & 5.2 \\
G12.50$-$0.22 & 36 & 3.55$^{+0.67}_{-0.75}$ & 13.05$^{+0.75}_{-0.67}$ & 3.55 & 5.2 \\
G14.33$-$0.57 & 19 & 1.99$^{+0.85}_{-0.95}$ & 14.48$^{+0.95}_{-0.85}$ & 1.99 & 2.1 \\
			& 20 & 2.04$^{+0.85}_{-0.94}$ & 14.43$^{+0.94}_{-0.85}$ & 2.04 & \\
G19.37$-$0.03 & 27 & 2.26$^{+0.88}_{-0.98}$ & 13.78$^{+0.97}_{-0.88}$ & 2.26 & 4.5 \\
G19.40$-$0.01 & 27 & 2.23$^{+0.88}_{-0.98}$ & 13.81$^{+0.97}_{-0.89}$ & 2.23 & 4.5 \\
G23.37$-$0.29 & 78 & 4.70$^{+0.90}_{-0.88}$ & 10.91$^{+0.88}_{-0.91}$ & 4.70 & 9.0 \\
			& 103 & 5.69$^{+1.20}_{-0.91}$ &  9.91$^{+0.92}_{-1.20}$ & & \\
			& 65 & 4.13$^{+0.88}_{-0.89}$ & 11.47$^{+0.90}_{-0.87}$ & & \\
G23.48$-$0.53 & 64 & 4.10$^{+0.88}_{-0.90}$ & 11.50$^{+0.89}_{-0.89}$ & 4.10 & 9.0 \\
			& 76 & 4.60$^{+0.91}_{-0.88}$ & 10.99$^{+0.88}_{-0.91}$ & & \\
G24.05$-$0.22 & 82 & 4.82$^{+0.96}_{-0.90}$ & 10.70$^{+0.91}_{-0.96}$ & 4.82 & 9.0 \\
G24.16$+$0.08 & 53 & 3.46$^{+0.91}_{-0.94}$ & 12.05$^{+0.94}_{-0.91}$ & 3.46 & 9.0 \\
			& 113 & 6.13$_{-1.03}$ & 9.38 & & \\
G25.99$-$0.06 & 90 & 5.15$^{+1.23}_{-0.99}$ & 10.13$^{+0.99}_{-1.23}$ & 5.15 & 14.0 \\
%G30.53$-$0.27 & 103 & 6.05 & 8.62 & 6.05 
G30.89$+$0.14 & 96 & 5.65$_{-1.38}$ & 8.93 & 5.65 & 8.5 \\
			& 40 & 2.62$^{+1.14}_{-1.15}$& 11.97$^{+1.15}_{-1.14}$ & & \\
			&108 & 6.65$_{-1.84}$ & 7.94 & & \\
G30.98$-$0.15 & 78 & 4.63$^{+1.63}_{-1.19}$ &  9.94$^{+1.19}_{-1.62}$ & 4.63 & 8.5 \\
G31.02$-$0.12 & 76 & 4.56$^{+1.56}_{-1.19}$ & 10.01$^{+1.19}_{-1.56}$ & 4.56 & 8.5 \\
			& 83 & 4.90$_{-1.23}$ & 9.67 & & \\
			& 92 & 5.41$_{-1.33}$ & 9.16 & & \\
G32.01$+$0.05 & 95 & 5.77$_{-1.51}$ & 8.64 & 5.77 & 8.5 \\
%G33.82$-$0.22 & 11 & 0.71$^{+1.25}$ & 13.41$^{+1.31}_{-1.25}$ & 0.71 & 3.7 \\
G34.63$-$1.03 & 14 & 0.84$^{+1.26}$ & 13.14$^{+1.32}_{-1.26}$ & 3.2 \\
G34.74$-$0.12 & 79 & 4.86$_{-1.45}$ & 9.11 & 4.86 & 3.7 \\
G34.78$-$0.80 & 44 & 2.80$^{+1.33}_{-1.26}$ & 11.17$^{+1.25}_{-1.33}$ & 2.80 & 3.2 \\
			& 37 & 2.41$^{+1.28}_{-1.26}$& 11.56$^{+1.26}_{-1.29}$ & & \\
G35.20$-$0.72 & 33 & 2.17$^{+1.29}_{-1.28}$ & 11.73$^{+1.27}_{-1.30}$ & 2.17 & 3.2 \\
G37.44$+$0.14 & 40 & 2.59$^{+1.47}_{-1.34}$ & 10.91$^{+1.34}_{-1.47}$ & 2.59 & 12.0 \\
			& 18 & 1.16$^{+1.34}$ & 12.34$^{+1.36}_{-1.34}$ & & \\
			& 86 & 5.90$_{-2.12}$ & 7.60 & & \\
G37.89$-$0.15 & 13 & 0.82$^{+1.35}$ & 12.60$^{+1.38}_{-1.35}$ & 0.82 & 12.0 \\
			& 65 & 4.16$_{-1.51}$ & 9.26 & & \\
			& 86 & 6.09$_{-2.31}$ & 7.33 & & \\
%G50.07$+$0.06 & 55 & 
\enddata
\end{deluxetable}
\end{center}

%\end{document}
\clearpage
%\documentclass[12pt,preprint]{aastex}
%\begin{document}
\begin{center}
\begin{deluxetable}{lcrclllcccc}
\label{corepropstab}
\tabletypesize{\scriptsize}
%\rotate
\tablecolumns{11}
\tablewidth{5.2in}
\tablecaption{Molecular Abundances and Cloud Masses}
\tablehead{
\colhead{Source} &
\colhead{Distance\tablenotemark{a}} &
\colhead{N(H$_{2}$)} &
%\colhead{$\Delta$N(H$_{2}$)} &
\colhead{} &
\multicolumn{3}{c}{Abundance Relative to H$_{2}$} &
\colhead{} &
\multicolumn{2}{c}{Cloud Mass} \\
\cline{5-7} \cline{9-10} \\
\colhead{} &
\colhead{(kpc)} &
\colhead{(10$^{21}$cm$^{-2}$)} &
%\colhead{(10$^{21}$cm$^{-2}$)} &
\colhead{} &
\colhead{N$_{2}$H$^{+}$} &
\colhead{C$^{18}$O} &
\colhead{CS} &
\colhead{} &
\colhead{(M$_{\odot}$)\tablenotemark{b}} &
\colhead{(M$_{\odot}$)\tablenotemark{c}}
}
\startdata
G05.85$-$0.23 &3.14 &4.4 $\pm$ 0.9 &&2.9E-10 &3.5E-7 &5.0E-10 &&2.6E+2 &2.5E+2\\
G06.26$-$0.51 &3.78 &13.5 $\pm$ 1.2 &&\nodata &1.3E-7 &5.6E-10 &&\nodata &6.2E+3\\
G09.16$+$0.06 &3.81 &4.4 $\pm$ 1.6 &&\nodata &3.4E-7 &4.2E-10 &&\nodata &3.3E+3\\
G09.21$-$0.22 &4.57 &1.9 $\pm$ 1.2 &&2.0E-9 &\nodata &3.2E-9 &&1.4E+3 &\nodata\\
G09.28$-$0.15 &4.48 &7.5 $\pm$ 1.2 &&5.2E-10 &\nodata &7.9E-10 &&3.4E+3 &\nodata \\
G09.86$-$0.04 &2.36 &9.8 $\pm$ 1.0 &&9.4E-11 &\nodata &7.2E-10 &&1.5E+3 &\nodata \\
%G09.88$-$0.11 &2.28 &2.6 $\pm$ 1.0 &&\nodata &\nodata &9.7E-10 &&\nodata &\nodata \\
G10.99$-$0.09 &3.32 &8.1 $\pm$ 0.9 &&5.5E-10 &3.0E-7 &5.7E-10 &&2.0E+3 &4.3E+3 \\
G12.22$+$0.14 &3.75 &2.8 $\pm$ 1.0 &&1.4E-9 &\nodata &3.0E-9 &&3.2E+2 &\nodata \\
G12.50$-$0.22 &3.55 &4.8 $\pm$ 1.7 &&6.5E-10 &3.6E-7 &1.4E-9 &&5.4E+2 &7.1E+3 \\
G14.33$-$0.57a &1.99 &4.1 $\pm$ 0.4 &&\nodata &5.7E-7 &1.4E-9 &&\nodata &8.3E+2 \\
G14.33$-$0.57b &2.05 &3.4 $\pm$ 1.1 &&4.4E-10 &5.7E-7 &1.1E-9 &&1.3E+3 &1.2E+3 \\
%G19.28-0.39 &3.86 &
G19.37$-$0.03 &2.26 &2.2 $\pm$ 1.5 &&1.7E-9 &\nodata &6.4E-9 &&2.8E+2 &\nodata \\
G19.40$-$0.01 &2.23 &4.9 $\pm$ 1.0 &&2.1E-10 &\nodata &7.7E-10 &&2.1E+3 &\nodata \\
G23.37$-$0.20 &4.70 &3.4 $\pm$ 1.1 &&9.0E-10 &1.2E-6 &3.1E-9 &&3.3E+3 &4.1E+3 \\
G23.48$-$0.53a &4.10 &7.8 $\pm$ 3.3 &&2.0E-10 &\nodata &5.4E-10 &&2.7E+3 &\nodata \\
G23.48$-$0.53b &4.02 &5.7 $\pm$ 1.4 &&2.3E-10 &\nodata &6.0E-10 &&2.0E+3 &\nodata \\
G24.05$-$0.22 &4.82 &2.7 $\pm$ 1.4 &&1.1E-9 &9.1E-7 &1.7E-9 &&4.0E+2 &2.1E+3 \\
G24.16$+$0.08 &3.46 &4.0 $\pm$ 1.4 &&\nodata &4.4E-7 &\nodata &&\nodata &2.6E+3 \\
G25.99$-$0.06 &5.15 &4.1 $\pm$ 1.5 &&2.1E-10 &\nodata  &1.2E-9 &&6.8E+2 &\nodata \\
%G30.49-0.39
%G30.53-0.27 &6.05 &3.13 &5.00 &&\nodata &\nodata &2.6E-9 &&\nodata &\nodata \\
G30.89$+$0.14 &5.65 &4.2 $\pm$ 1.0 &&3.2E-10 &\nodata &6.2E-10 &&1.1E+4 &\nodata \\
G30.98$-$0.15 &4.63 &7.3 $\pm$ 2.7 &&6.6E-10 &\nodata &1.4E-9 &&1.9E+3 &\nodata \\
G31.02$-$0.12 &4.56 &4.3 $\pm$ 0.9 &&3.9E-10 &3.1E-7 &7.1E-10 &&2.6E+3 &4.7E+3 \\
G32.01$+$0.05 &5.77 &7.3 $\pm$ 2.3 &&1.0E-9 &6.5E-7 &2.6E-9 &&8.7E+3 &1.3E+4 \\
%G33.82$-$0.22 &0.71 &2.0 $\pm$ 0.5 &&5.4E-10 &\nodata &1.1E-9 &&3.3E+0 &\nodata \\
G34.63$-$1.03 &0.84 &3.6 $\pm$ 1.0 &&3.4E-10 &1.7E-7 &\nodata &&5.5E+1 &6.0E+1 \\
G34.74$-$0.12 &4.86 &5.1 $\pm$ 1.5 &&4.0E-10 &7.3E-7 &1.0E-9 &&8.7E+2 &2.4E+3 \\
G34.78$-$0.80 &2.80 &6.1 $\pm$ 3.8 &&5.7E-11 &3.9E-7 &1.4E-9 &&\nodata &2.9E+3 \\
G35.20$-$0.72 &2.17 &3.5 $\pm$ 2.5 &&9.2E-10 &6.2E-7 &2.5E-9 &&1.0E+3 &1.5E+3 \\
G37.44$+$0.14a &2.59 &3.0 $\pm$ 1.5 &&\nodata &4.0E-7 &1.2E-9 &&\nodata &8.9E+1 \\
G37.44$+$0.14b &1.16 &2.6 $\pm$ 0.9 &&2.6E-10 &2.1E-7 &4.5E-10 &&3.2E+1 &1.1E+2 \\
G37.89$-$0.15 &0.82 &3.2 $\pm$ 1.6 &&1.5E-10 &2.1E-7 &5.2E-10 &&3.7E+0 &5.4E+1\\
%G43.64-0.82
%G50.07+0.06
\enddata
%\tablenotetext{a}{Distances found using the rotation curve of Fich, Blitz, \& Stark, 1989, ApJ, 342, 272}
\tablenotetext{a}{Distances from rotation curve of Fich, Blitz, \& Stark (1989), assuming absorbing cloud lies at near distance.}
%\tablenotetext{b}{Values given as e.g. 1.2E+4 to be taken as 1.2 $\times$ 10$^{4}$}
\tablenotetext{b}{from N$_{2}$H$^{+}$}
\tablenotetext{c}{from C$^{18}$O}
\end{deluxetable}
\end{center}
%\end{document}
\clearpage

\begin{figure}
\caption[singlespace]{
The color image is the 8 $\mu$m MSX, and the colorbar indicates the emission intensity scale in W m$^{-2}$ sr$^{-1}$.  The contours represent the molecular line integrated intensity, for which the levels are listed for each object.  The integrated emission was calculated within a 6 -- 10 km~s$^{-1}$ window for C$^{18}$O and CS.  For N$_2$H$^+$ the window was 15 -- 20 km~s$^{-1}$ with some gaps accounting for the clustering of the hyperfine components.
\\\textit{G05.85$-$0.23} - blue contours show emission at 17km~s$^{-1}$; green
contours: 9km~s$^{-1}$.  Levels: 1,2 K km~s$^{-1}$ for all plots.
\\\textit{G06.26$-$0.51} - blue contours : 23km~s$^{-1}$; green contours : 17km
s$^{-1}$. Levels: N$_{2}$H$^{+}$ : 0.5, 1 K km~s$^{-1}$.  CS : 1,2,3,4 K km~s$^{-1}$. 
C$^{18}$O : 1,2,3 K km~s$^{-1}$ for both velocities.
\\\textit{G09.16$+$0.06} - blue contours : 31km~s$^{-1}$.  Levels:  N$_{2}$H$^{+}$ : 1,2 K km~s$^{-1}$.  CS : 0.5,1,1.5 K km~s$^{-1}$.  C$^{18}$O : 1,2,3 K km~s$^{-1}$. 
\\\textit{G09.21$-$0.22} - blue contours : 43km~s$^{-1}$. 
Levels: N$_{2}$H$^{+}$ : 2,4,6,8,10,12 K km~s$^{-1}$.  CS: 1,2,3 K km~s$^{-1}$.
\\\textit{G09.28 $-$0.15} - blue contours : 42km~s$^{-1}$.  Levels:  N$_{2}$H$^{+}$ :
1.5,3,4.5,6,7.5 K km~s$^{-1}$.  CS: 0.5,1,1.5,2,3 K km~s$^{-1}$.
\\\textit{G09.86$-$0.04} - blue contours : 18km~s$^{-1}$.  Levels:  N$_{2}$H$^{+}$ :
1.5,3,4.5,6,7.5 K km~s$^{-1}$.  CS: 0.5,1,1.5,2,3 K km~s$^{-1}$. 
\\\textit{G09.88 $-$0.11} - blue contours : 17km~s$^{-1}$. 
Levels: N$_{2}$H$^{+}$ : 1,2 K km~s$^{-1}$.  CS: 0.5,1,1.5,2,3 K km~s$^{-1}$.
\\\textit{G10.59$-$0.31} - blue contours : 17km~s$^{-1}$.  Levels: 3,5,7,9,11 K km
s$^{-1}$ for both molecules.
\\\textit{G10.70$-$0.33} - blue contours : 0km~s$^{-1}$; green contours : 33km~s$^{-1}$. 
Levels: 2,3,4,5 K km~s$^{-1}$ for both molecules/velocities. 
\\\textit{G10.99$-$0.09} - blue contours : 30km~s$^{-1}$. 
Levels: N$_{2}$H$^{+}$ : 2,4,6,8,10 K km~s$^{-1}$.  CS: 1,1.5,2 K km~s$^{-1}$.  C$^{18}$O :
2,3,4,5 K km~s$^{-1}$.
\\\textit{G12.22$+$0.14} - blue contours : 40 km~s$^{-1}$.  Levels:  1,2,3,4,5 K
km~s$^{-1}$ for each molecule.
\\\textit{G12.50$-$0.22} - blue contours : 36 km~s$^{-1}$.  Levels:  N$_{2}$H$^{+}$ :
2,4,6,8,10 K km~s$^{-1}$.  CS: 1,2,3,4 K km~s$^{-1}$.  C$^{18}$O : 1,2,3 K km~s$^{-1}$. 
\\\textit{G14.33$-$0.57} - blue contours : 19km~s$^{-1}$ (solid (a)), 20km s${-1}$ (dashed (b))  Levels: N$_{2}$H$^{+}$:  2,4,6,8,10 K km~s$^{-1}$.  CS : 2,3,4,5,6,7 K km~s$^{-1}$.  C$^{18}$O : 4,5,6,7,8,9 K km~s$^{-1}$ (a); 2,3,4 K km~s$^{-1}$ (b).
\\\textit{G19.37$-$0.03} - blue contours : 27km~s$^{-1}$.  Levels: N$_{2}$H$^{+}$ :
1,3,5,7,9,11 K km~s$^{-1}$.  CS : 1,2.5,4,5.5,7 K km~s$^{-1}$.
\\\textit{G19.40$-$0.01} - blue contours : 27km~s$^{-1}$.  Levels: N$_{2}$H$^{+}$ :
1,3,5,7,9,11 K km~s$^{-1}$.  CS : 1,2.5,4,5.5,7 K km~s$^{-1}$. 
\\\textit{G23.37$-$0.29} - blue contours : 78km~s$^{-1}$; green
contours : 103km~s$^{-1}$; white contours : 65km~s$^{-1}$.  Levels:
N$_{2}$H$^{+}$ : 3,6,9,12,15,18 K km~s$^{-1}$ (78km~s$^{-1}$ component); 4,6,8,10 K km~s$^{-1}$ (103km~s$^{-1}$).  CS : 3,4,5,6,7 K km~s$^{-1}$ (78km~s$^{-1}$);
3,4,5,6 K km~s$^{-1}$ (103km~s$^{-1}$); 3,4,5,6,9 K km~s$^{-1}$ (65km~s$^{-1}$).  C$^{18}$O : 5,7,9,11 K km~s$^{-1}$ (78km~s$^{-1}$); 3,6,9,12,15,18,21,24 K km~s$^{-1}$ (103km~s$^{-1}$); 3,6,9 K km~s$^{-1}$ (65km~s$^{-1}$).
\\\textit{G23.48 $-$0.53} - blue contours : 64km~s$^{-1}$; green contours : 76km~s$^{-1}$. 
Levels:  1,1.5,2,2.5 K km~s$^{-1}$ both molecules/velocities.
\\\textit{G24.05$-$0.22} - blue contours : 82km~s$^{-1}$.  Levels: N$_{2}$H$^{+}$ :
1,2,3 K km~s$^{-1}$.  CS : 1,2,3,4 K km~s$^{-1}$.  C$^{18}$O : 1,2,3,4,5 K km
s$^{-1}$. 
\\\textit{G24.16$+$0.08} - blue contours : 53km~s$^{-1}$; green
contours : 113km~s$^{-1}$.  Levels:  N$_{2}$H$^{+}$ : 1,2,3 K km~s$^{-1}$ (113km
s$^{-1}$ component).  CS : 1,2,3 K km~s$^{-1}$ in both velocities.  C$^{18}$O : 1,2,3 K km~s$^{-1}$ (53km~s$^{-1}$); 2,4,6 K km~s$^{-1}$ (113km~s$^{-1}$).
\\\textit{G25.99$-$0.06} - blue contours : 90km~s$^{-1}$.  Levels: N$_{2}$H$^{+}$ :
1,2,3 K km~s$^{-1}$.  CS : 1,2 K km~s$^{-1}$.
\\\textit{G30.14$-$0.07} - blue contours : 87km~s$^{-1}$.  Levels: 0.5 for both molecules. 
\\\textit{G30.53$-$0.27} - blue contours : 103km~s$^{-1}$. 
Levels: N$_{2}$H$^{+}$ : 0.5 K km~s$^{-1}$.  CS : 1,1.5,2,2.5 K km~s$^{-1}$.
\\\textit{G30.89$+$0.14} - blue contours : 96km~s$^{-1}$; green contours : 40km~s$^{-1}$; white contours : 108km~s$^{-1}$.  Levels: N$_{2}$H$^{+}$ : 2,3,4 K km~s$^{-1}$ (96km~s$^{-1}$ component); 2,4,6,8,10 K km~s$^{-1}$ (40km~s$^{-1}$); 2,4,6,8,10 K km~s$^{-1}$ (108km~s$^{-1}$). CS : 3,4,5,6,7 K km~s$^{-1}$ (96km~s$^{-1}$); 2,3,4 K km~s$^{-1}$ (40km~s$^{-1}$ and 108km~s$^{-1}$ components).
\\\textit{G30.98 $-$0.15} - blue contours : 78km~s$^{-1}$.  Levels: N$_{2}$H$^{+}$ :
1,2,4,6,8 K km~s$^{-1}$.  CS : 1,2,3,4 K km~s$^{-1}$.  
\\\textit{G31.02$-$0.12} - blue contours : 76km~s$^{-1}$; green
contours : 83km~s$^{-1}$; white contours : 92km~s$^{-1}$.  Levels:
N$_{2}$H$^{+}$ : 1,2,3,4 K km~s$^{-1}$ (76km~s$^{-1}$ and 83km~s$^{-1}$ components);
2,4,6,8,10 K km~s$^{-1}$ (92km~s$^{-1}$).  CS : 1,2,3,4,5,6,7 K km~s$^{-1}$
(76km~s$^{-1}$); 1,2,3,4,5,6 K km~s$^{-1}$ (83km~s$^{-1}$); 1,2,3,4,5,6,9 K km~s$^{-1}$ (92km~s$^{-1}$).  C$^{18}$O : 1,2,3,4,5,6,7,8,9 K km~s$^{-1}$
(76km~s$^{-1}$); 1,2,3,6,9,12,15,18,21,24 K km~s$^{-1}$ (83km~s$^{-1}$);
1,2,3 K km~s$^{-1}$ (92km~s$^{-1}$)
\\\textit{G32.01$+$0.05} - blue contours : 95km~s$^{-1}$.  Levels: N$_{2}$H$^{+}$ :
3,6,9,12,15,18,21,24 K km~s$^{-1}$.  CS : 2,4,6,8,10 K km~s$^{-1}$.  C$^{18}$O : 2,4,6,8,10 K km~s$^{-1}$.
\\\textit{G33.82$-$0.22} - blue contours : 11km~s$^{-1}$.  Levels: N$_{2}$H$^{+}$ :
1,2,3 K km~s$^{-1}$.  CS : 0.5,1 K km~s$^{-1}$. 
\\\textit{G34.63$-$1.03} - blue contours : 14km~s$^{-1}$. 
Levels: N$_{2}$H$^{+}$ : 1,2,3 K km~s$^{-1}$.  CS : 0.5,1.0 K km~s$^{-1}$.  C$^{18}$O :
0.5,1,1.5 K km~s$^{-1}$.
\\\textit{G34.74$-$0.12} - blue contours : 79km~s$^{-1}$.  Levels: N$_{2}$H$^{+}$ :
1,2,3 K km~s$^{-1}$.  CS : 1,2,3,4 K km~s$^{-1}$.  C$^{18}$O : 1,2,3,4,5 K km
s$^{-1}$.
\\\textit{G34.78 $-$0.80} - blue contours : 44km~s$^{-1}$; green contours : 37km~s$^{-1}$. 
Levels: N$_{2}$H$^{+}$ : 1,2,3,4 K km~s$^{-1}$ for both velocities.  CS : 1,2,3 K
km~s$^{-1}$ for both velocities.  C$^{18}$O : 1,2,3 K km~s$^{-1}$ (44km~s$^{-1}$
component); 1,2,3,4,5,6 K km~s$^{-1}$ (37km~s$^{-1}$). 
\\\textit{G35.20$-$0.72} - blue contours : 33km~s$^{-1}$. 
Levels: N$_{2}$H$^{+}$ : 2,4,6,8,10,12,14 K km~s$^{-1}$.  CS : 2,4,6,8,10,12 K km
s$^{-1}$.  C$^{18}$O : 2,4,6,8,10,12,14 K km~s$^{-1}$.
\\\textit{G37.44$+$0.14} - blue contours : 40 km~s$^{-1}$; green contours : 18km~s$^{-1}$; white contours : 86km~s$^{-1}$.  Levels: 0.5,1,2,3 K km~s$^{-1}$ for all molecules/velocities.
\\\textit{G37.89$-$0.15} - blue contours : 13km~s$^{-1}$; green contours : 65km~s$^{-1}$; white contours : 86km~s$^{-1}$.  Levels: N$_{2}$H$^{+}$ : 0.5.1,2,3 K km~s$^{-1}$ (13km~s$^{-1}$ component).  CS : 0.5.1,2,3 K km~s$^{-1}$ (13km~s$^{-1}$).  C$^{18}$O : 0.5,1,2,3 K km~s$^{-1}$ (13km~s$^{-1}$
and 65km~s$^{-1}$); 2,3 K km~s$^{-1}$ (86km~s$^{-1}$). 
\\\textit{G50.07$+$0.06} - blue contours 55km~s$^{-1}$. Levels: 0.5,1 K km~s$^{-1}$ for both molecules.
}
\label{fig:maps}
\end{figure}
\clearpage
%\begin{comment}
%\epsscale{.8}
\plotone{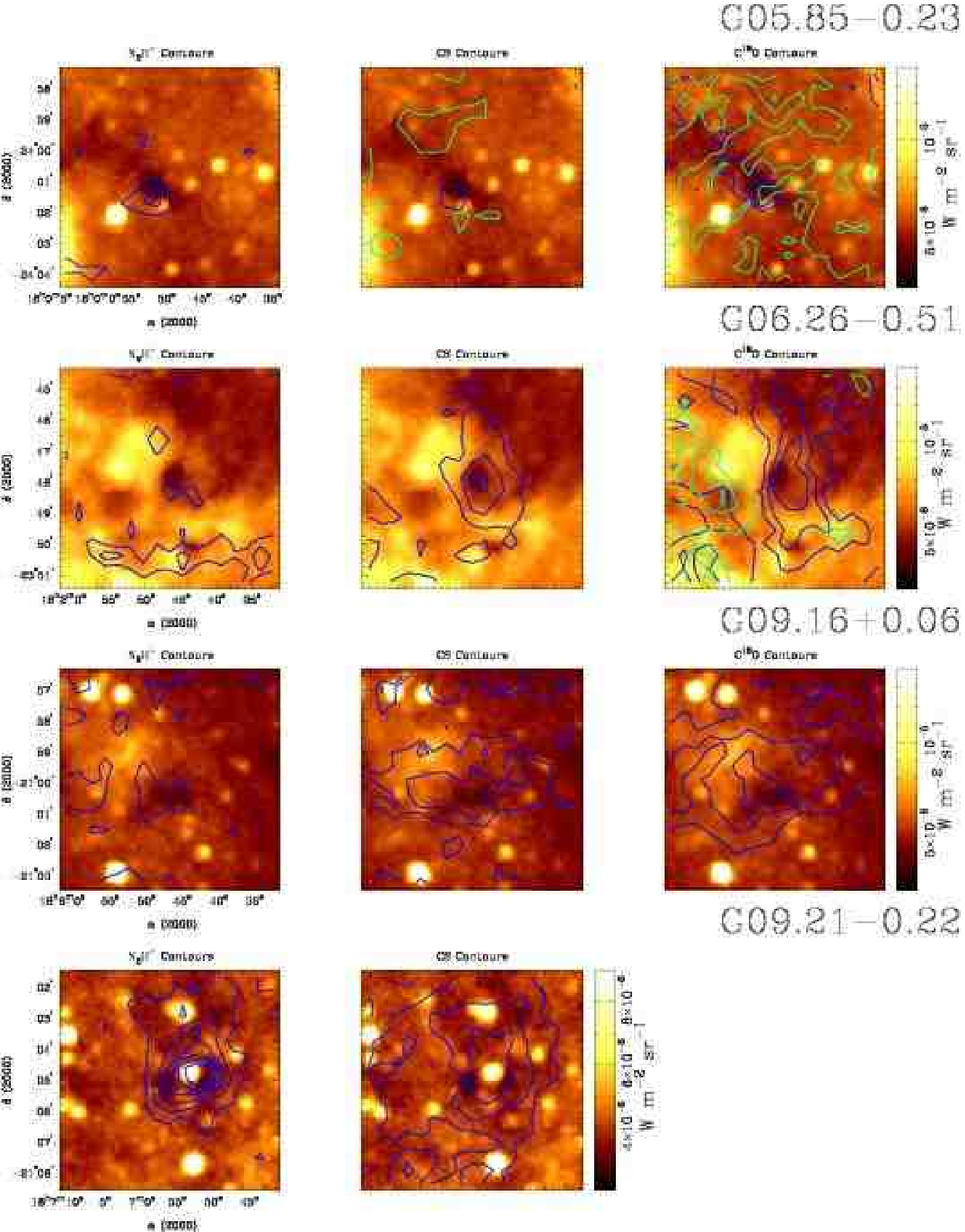}
\centerline{Fig. 1.}
\clearpage
\plotone{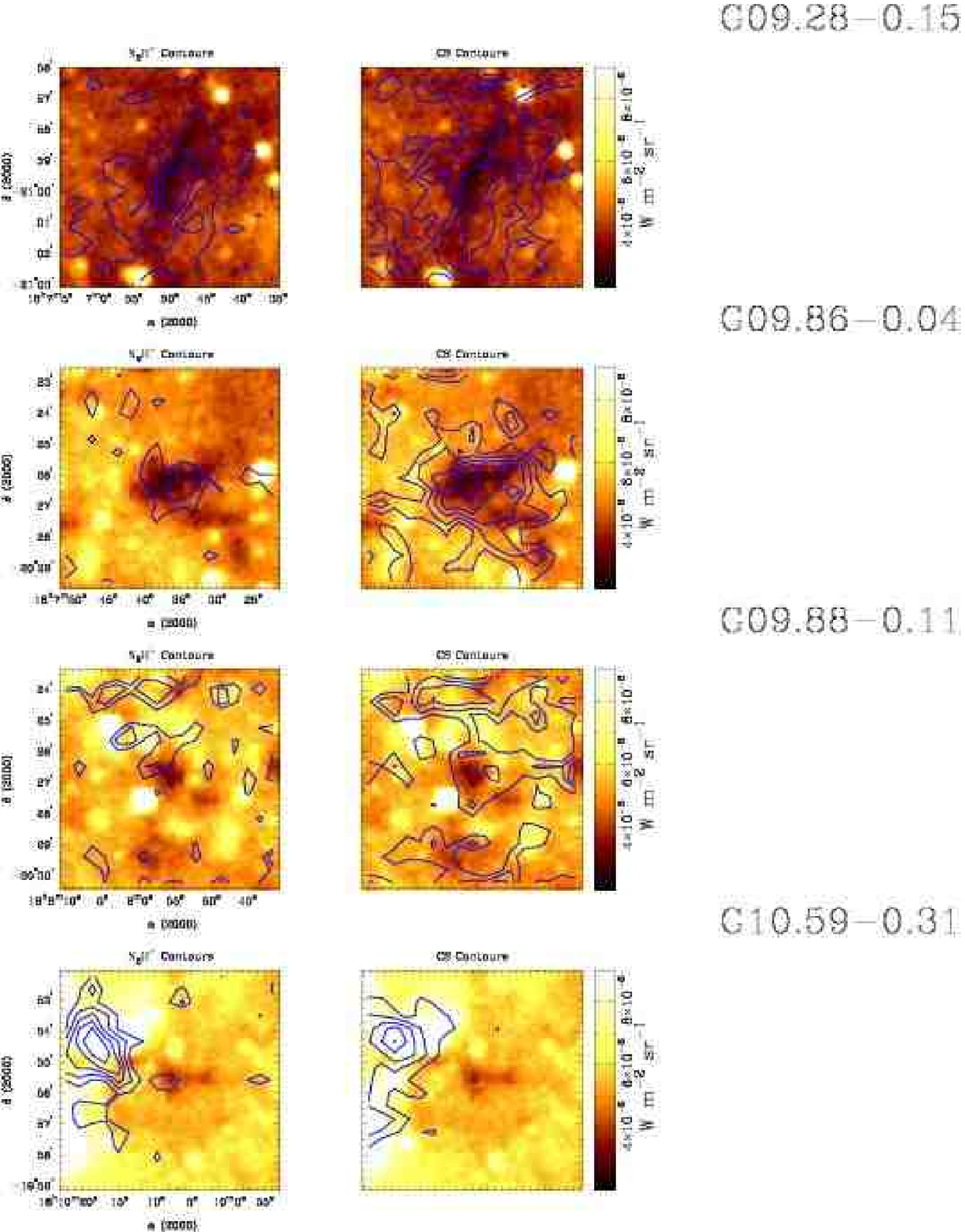}
\centerline{Fig. 1. --- Continued}
\clearpage
\plotone{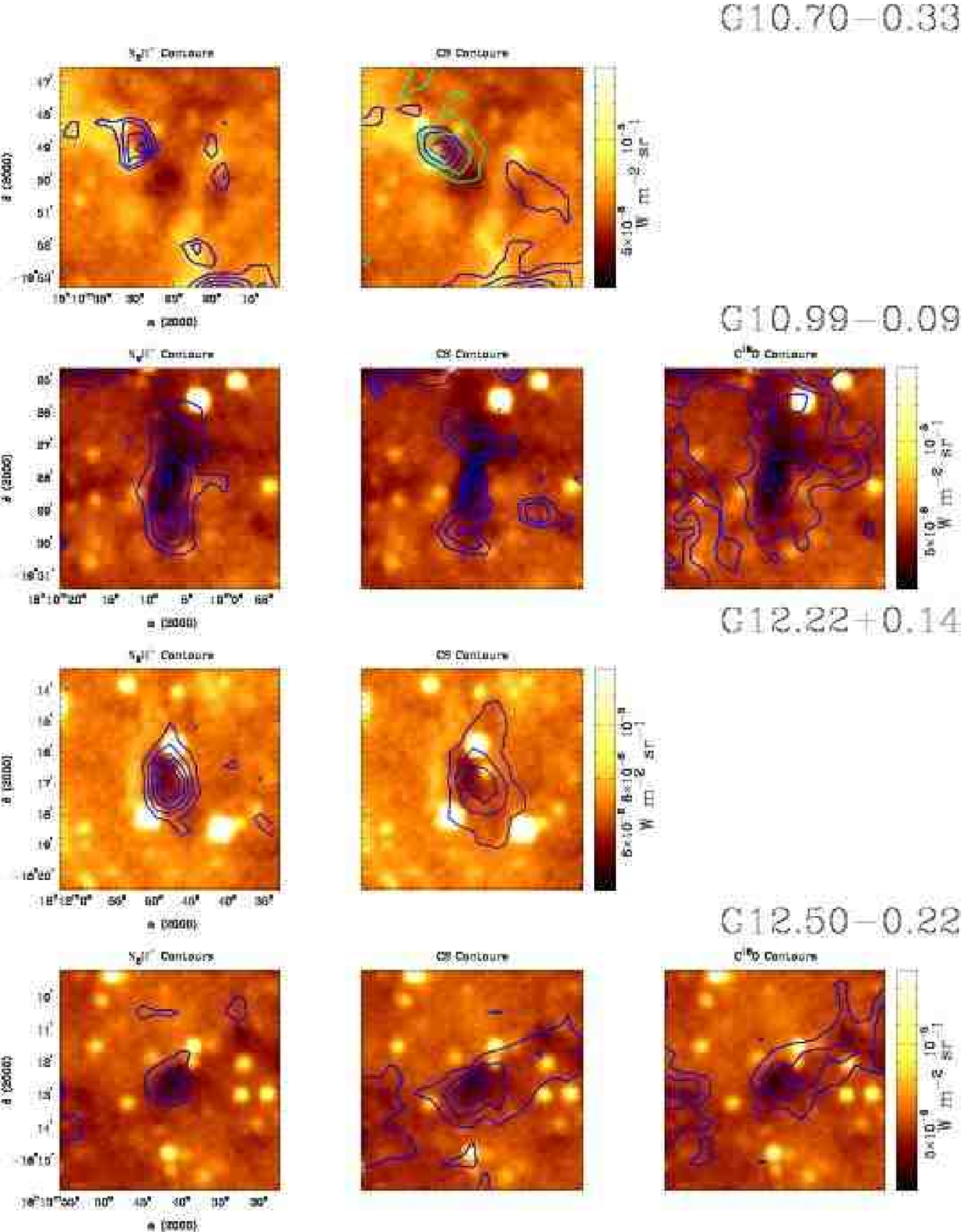}
\centerline{Fig. 1 --- Continued}
\clearpage
\plotone{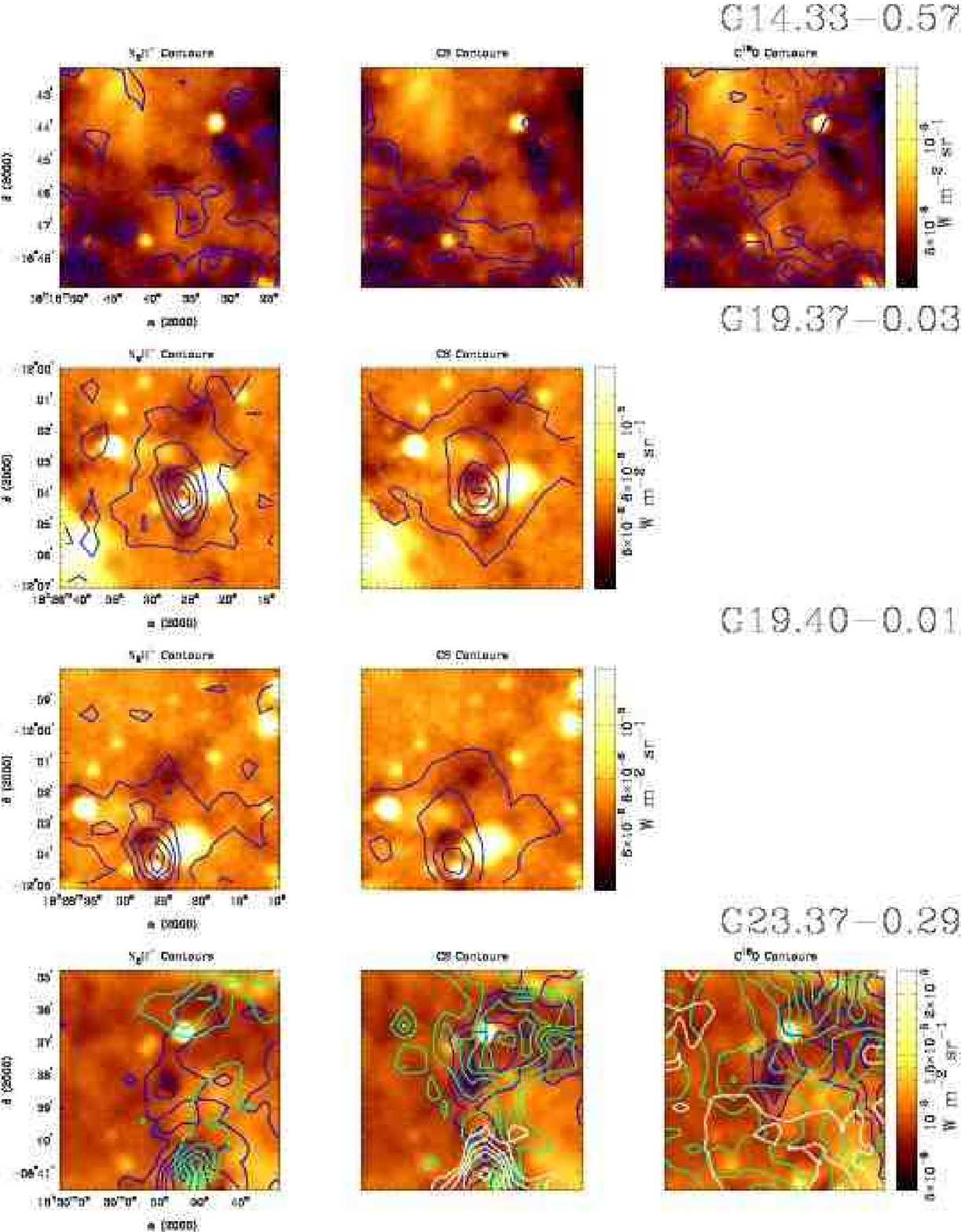}
\centerline{Fig. 1 --- Continued}
\clearpage
\plotone{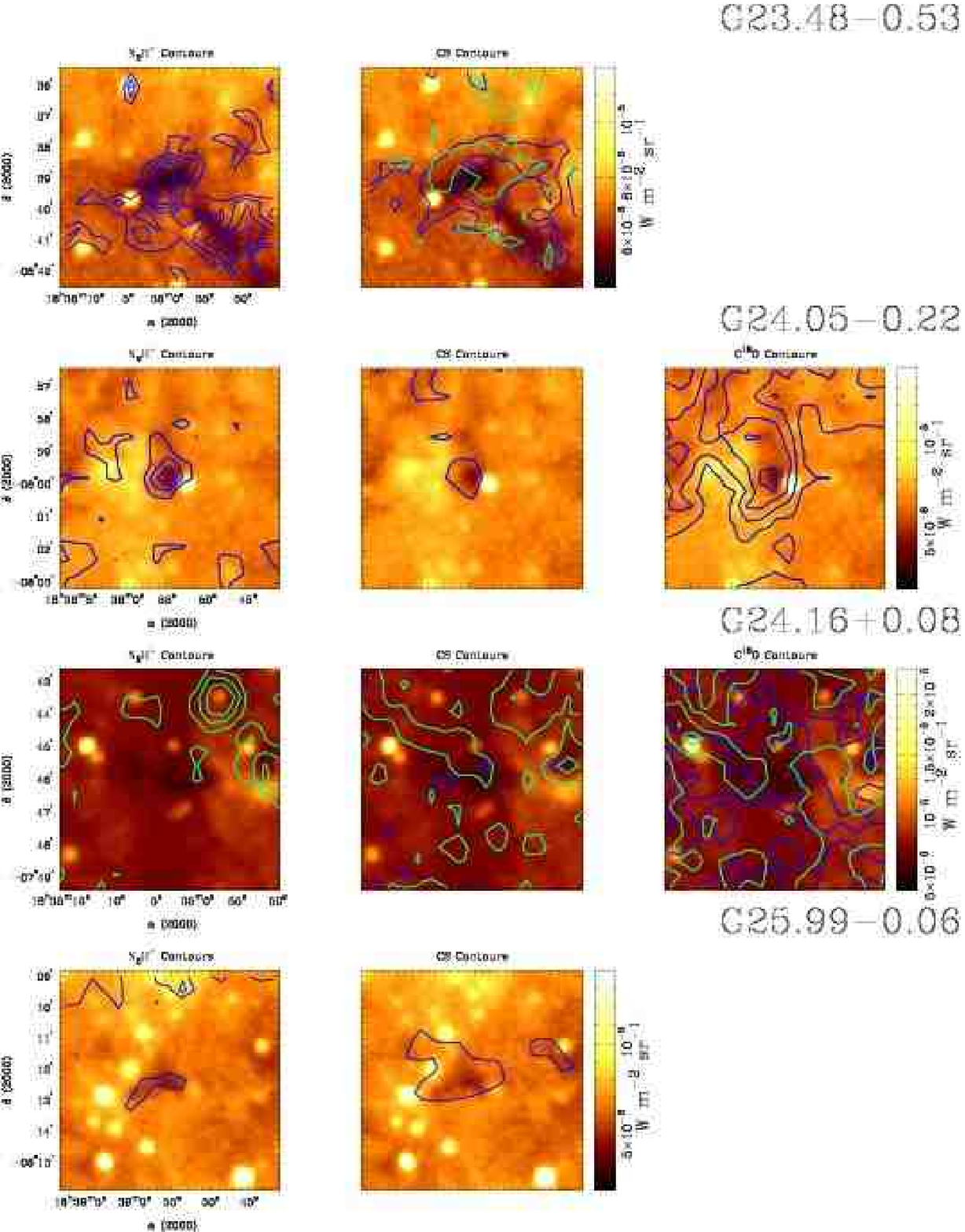}
\centerline{Fig. 1 --- Continued}
\clearpage
\plotone{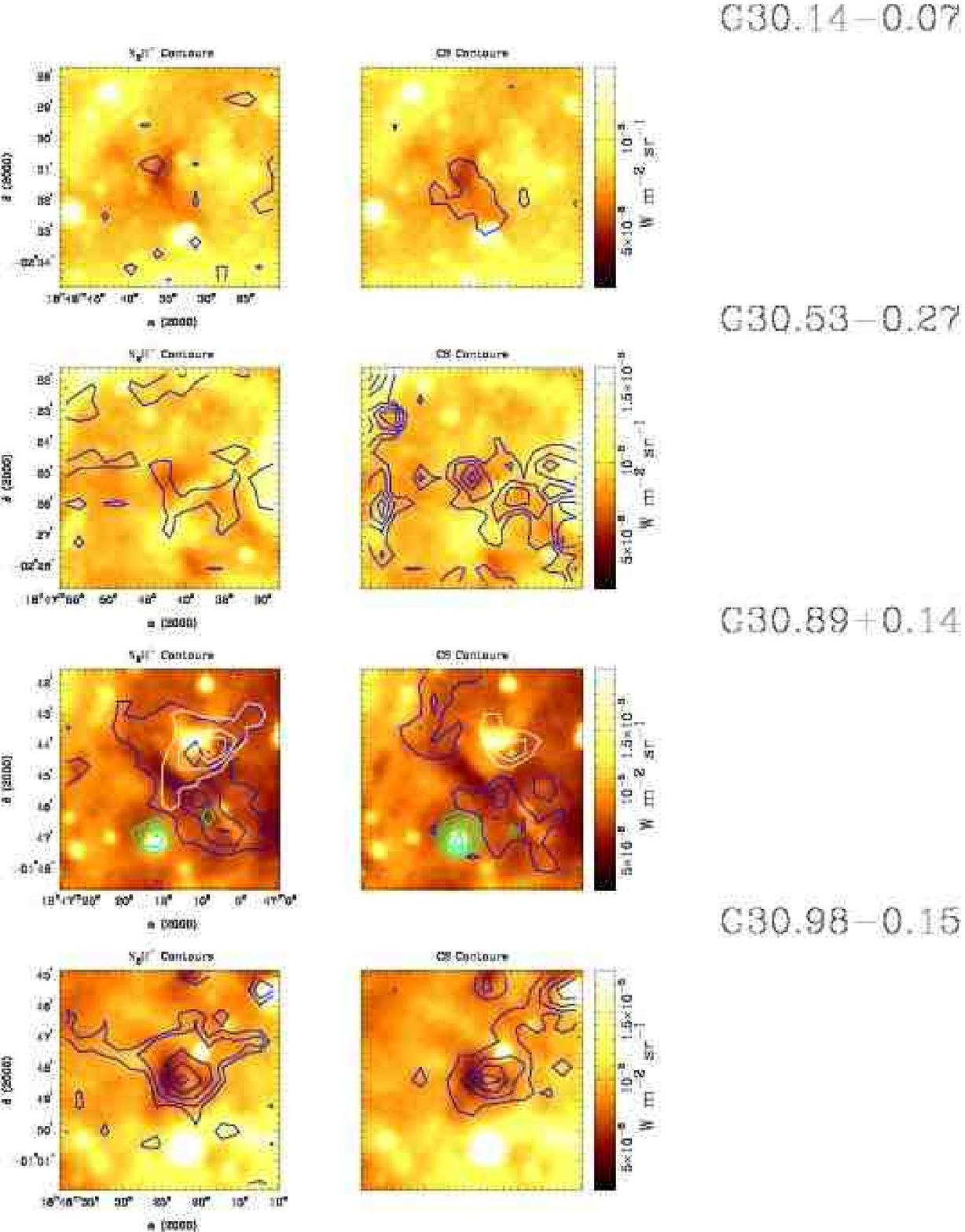}
\centerline{Fig. 1 --- Continued}
\clearpage
\plotone{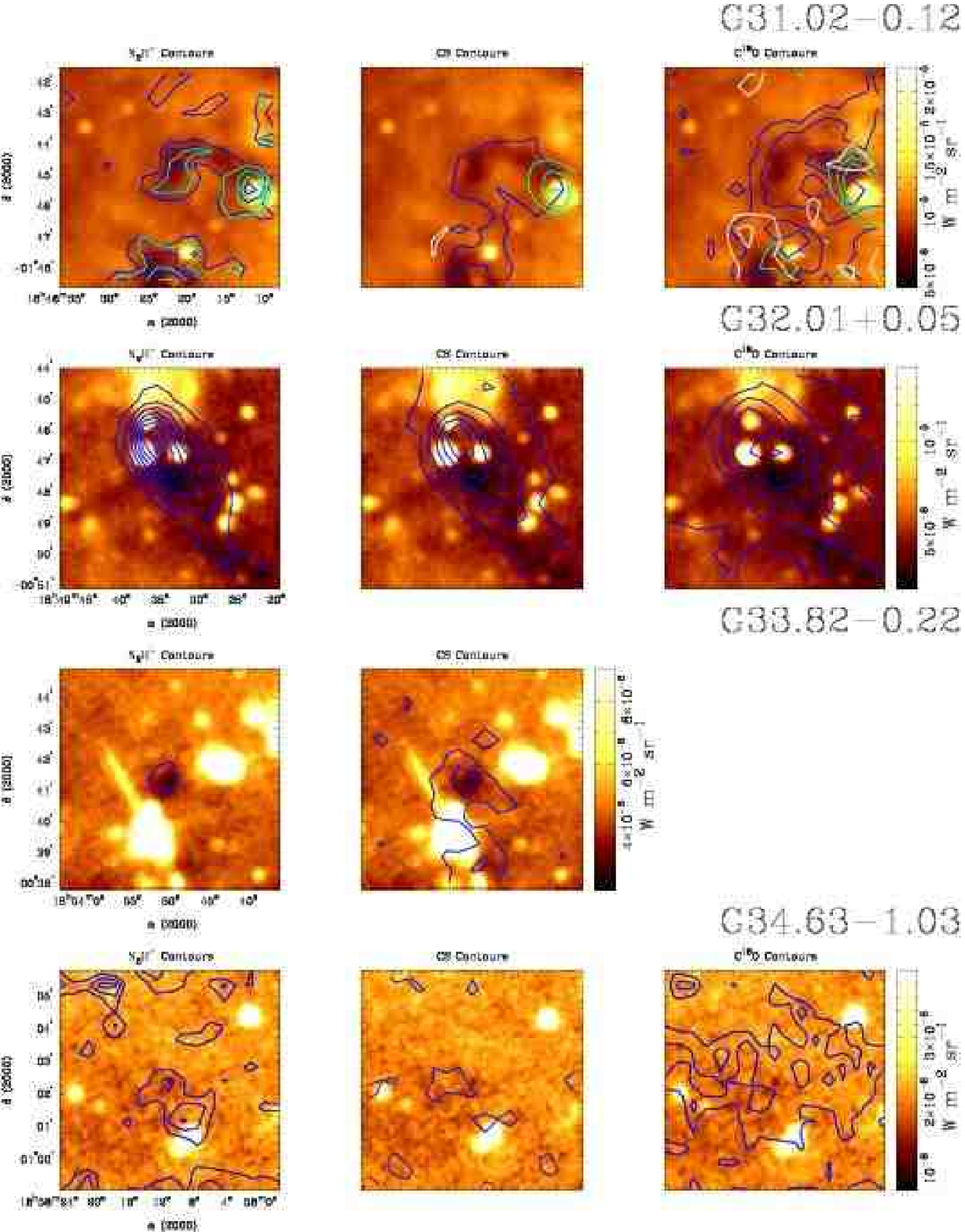}
\centerline{Fig. 1 --- Continued}
\clearpage
\plotone{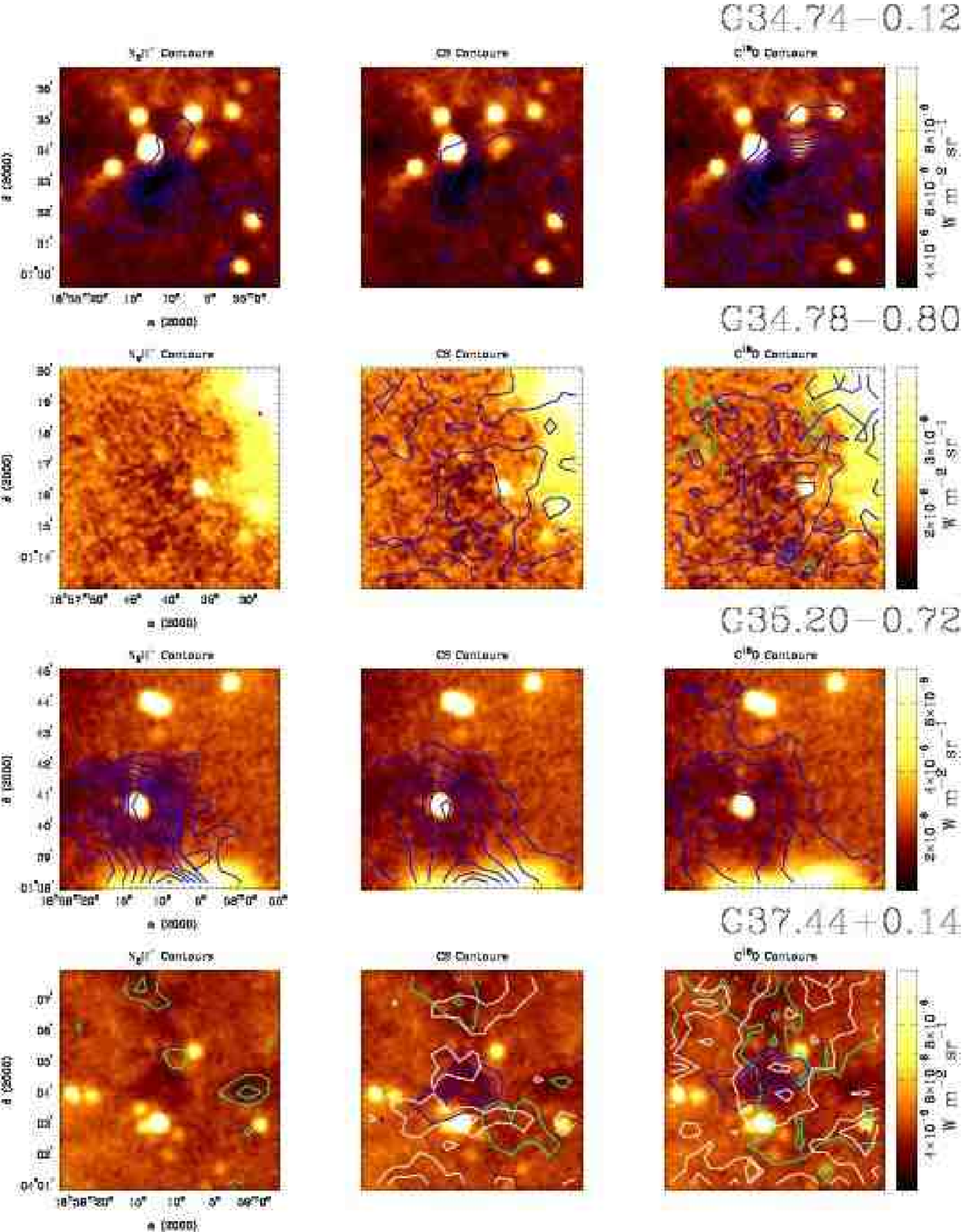}
\centerline{Fig. 1 --- Continued}
\clearpage
\plotone{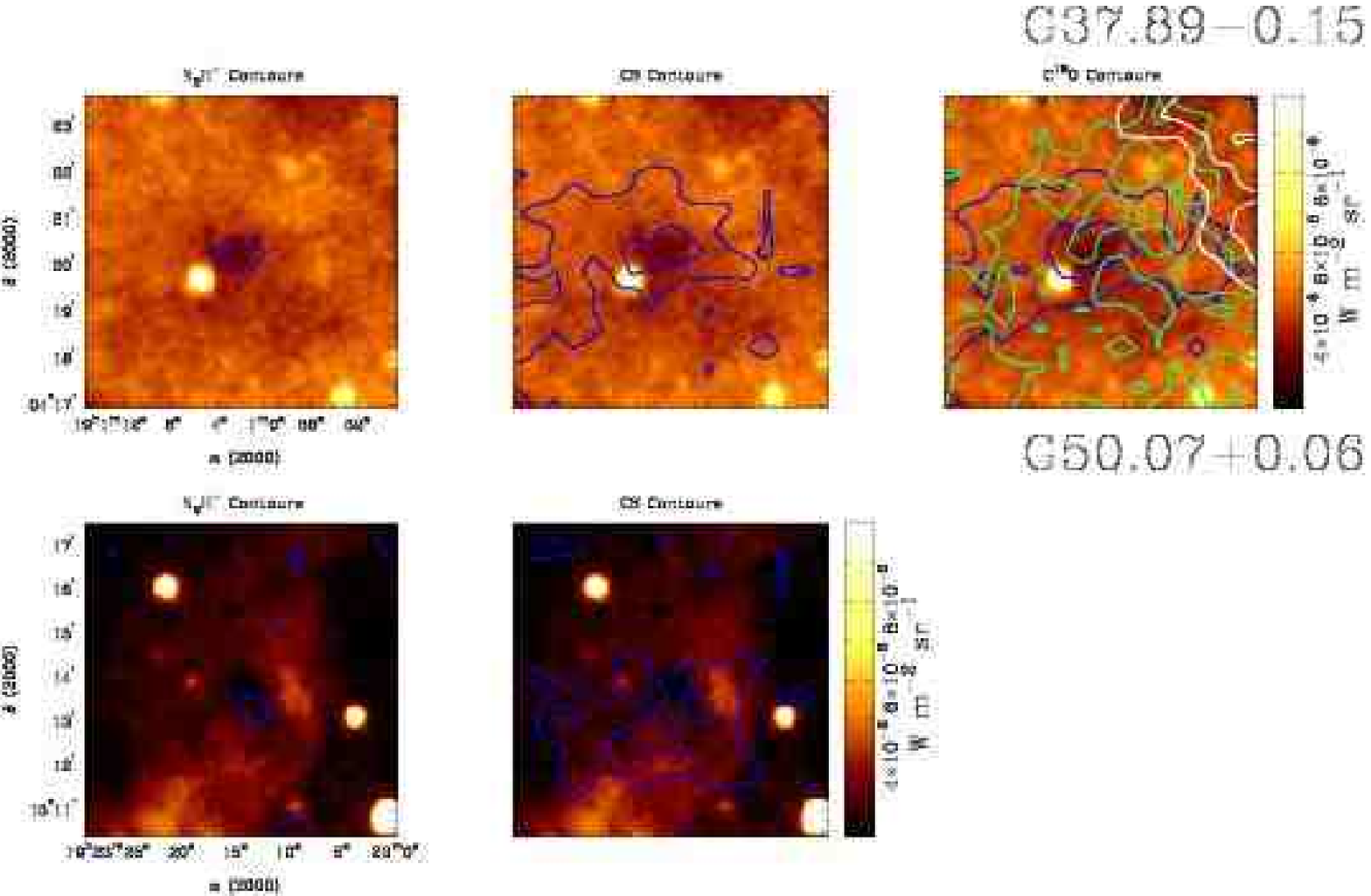}
\centerline{Fig. 1 --- Continued}
\clearpage
%\end{comment}
%\renewcommand{\thefigure}{\arabic{figure}}

\begin{figure}
\epsscale{1}
\begin{center}
\plottwo{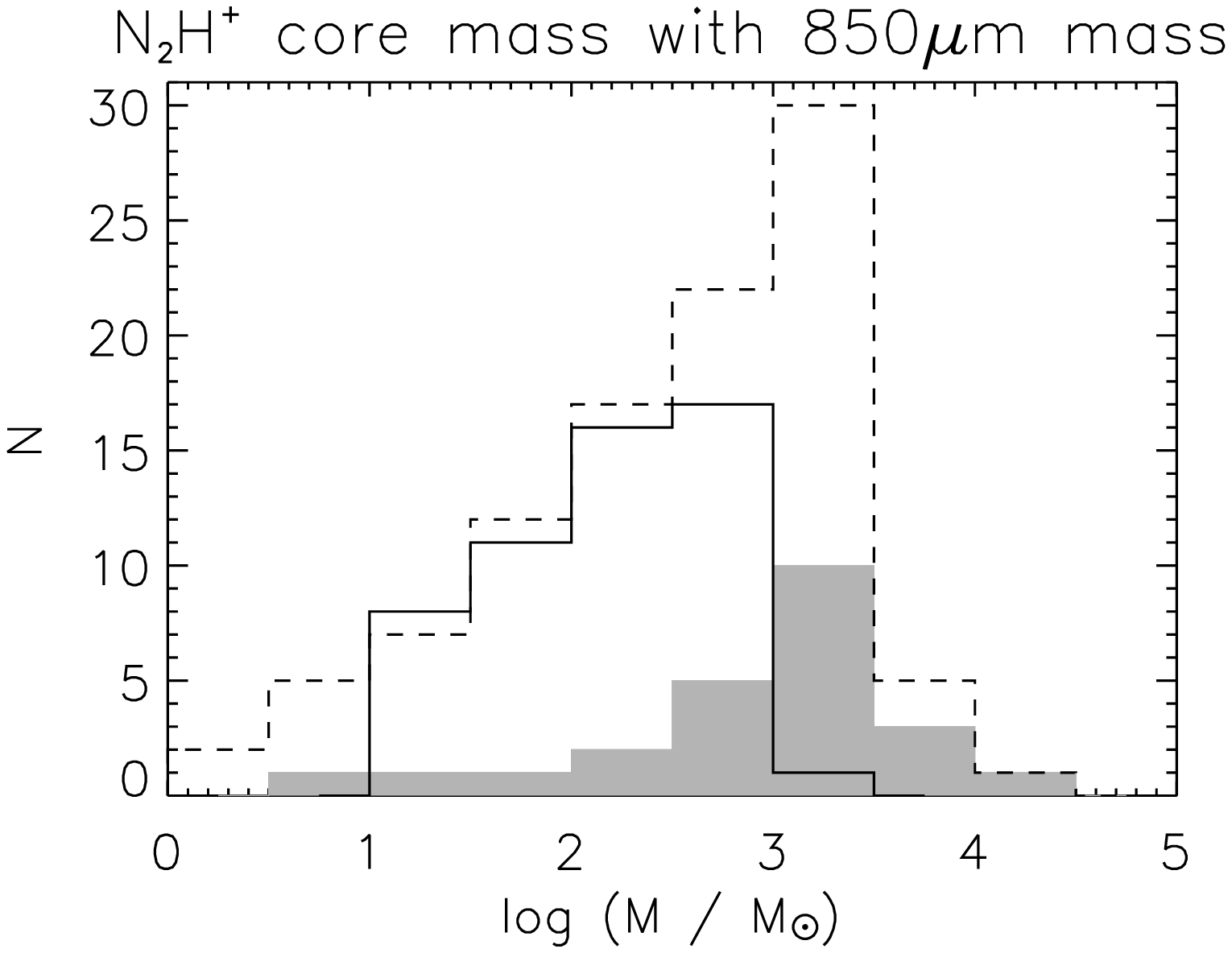}{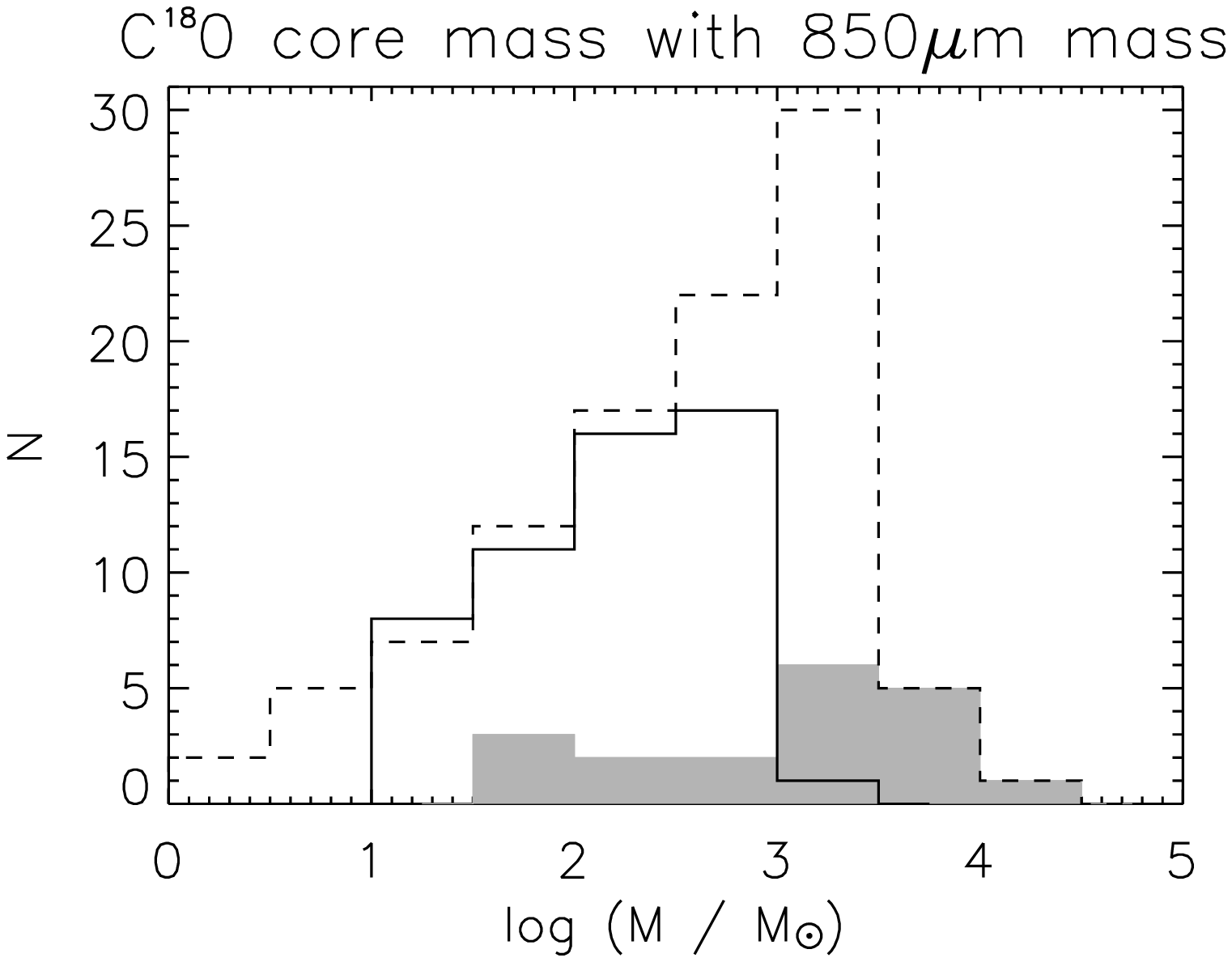}
\end{center}
\caption{Histogram of the mass distribution for our sample of IR dark cores as derived from the N$_2$H$^+$ data (left panel) and the C$^{18}$O data (right panel).  The grey histograms show the mass distribution for our sample of cores whereas the solid line shows the distribution
of masses of the Williams et al. (2004) cores (derived from 850$\mu$m data) based on the near galactic distances.  The dashed line represents the masses based on the far galactic distances.}
\label{fig:mass}
\end{figure}

\begin{figure}
\begin{center}
\includegraphics[scale=0.7]{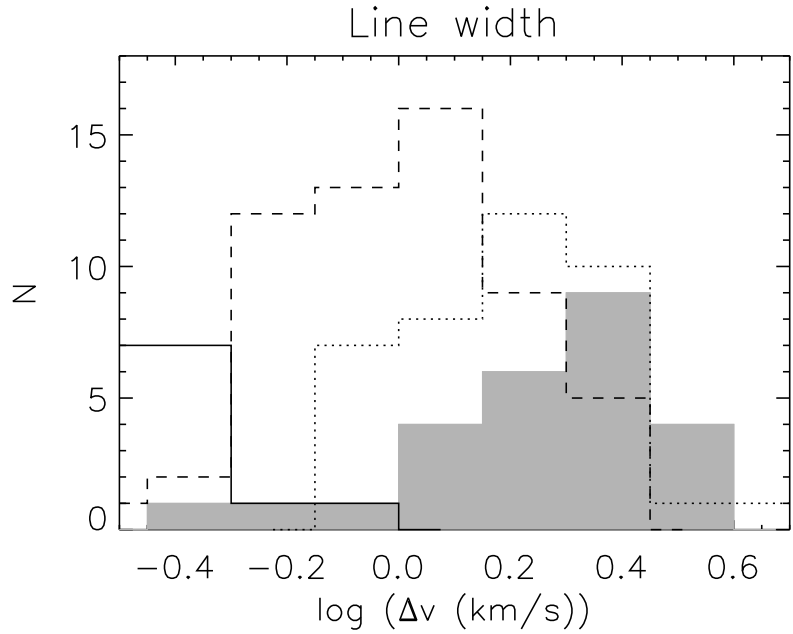}
\end{center}
\caption{Histogram of the N$_2$H$^+$ (solid line, Caselli et al. 2002; grey, this sample) and NH$_3$ (dashed line, Harju et al. 1993; dotted line, Molinari et al. 1996) linewidths.}
\label{fig:vel}
\end{figure}

\end{document}